\documentclass[longbibliography,aps,reprint,groupedaddress,superscriptaddress]{revtex4-2}

\usepackage[pdftex]{graphicx}
\usepackage{dcolumn}
\usepackage{bm}
\usepackage{amsmath}
\usepackage{latexsym}
\usepackage{epstopdf}
\usepackage{xcolor}
\usepackage{lipsum}
\usepackage{xargs}

\usepackage{lipsum}
\usepackage{float}

\usepackage{hyperref}
\definecolor{bluemy}{rgb}{0.157, 0.173, 0.569}
\hypersetup{colorlinks=true, linkcolor=bluemy, citecolor=bluemy, urlcolor=bluemy,anchorcolor=bluemy}

\newcommand{\rfig}[1]{Fig.~\ref{#1}}
\newcommand{\rfigs}[1]{Figs.~\ref{#1}}

\newcommand{\Rmnum}[1]{\expandafter\@slowromancap\romannumeral #1@}
 
\DeclareMathSizes{10}{9}{7}{5}
\makeindex

\begin{document}
\title{Synthetic multi-dimensional Aharonov-Bohm cages in Fock state lattices}
\author{Jiajian Zhang}
\thanks{These authors contributed equally to this work.}
	\affiliation{Shenzhen Institute for Quantum Science and Engineering, Southern University of Science and Technology, Shenzhen 518055, China}
	\affiliation{International Quantum Academy, Shenzhen 518048, China}
	\affiliation{Guangdong Provincial Key Laboratory of Quantum Science and Engineering, Southern University of Science and Technology, Shenzhen 518055, China}
	
	\author{Wenhui Huang}
	\thanks{These authors contributed equally to this work.}
	\affiliation{Shenzhen Institute for Quantum Science and Engineering, Southern University of Science and Technology, Shenzhen 518055, China}
	\affiliation{International Quantum Academy, Shenzhen 518048, China}
	\affiliation{Guangdong Provincial Key Laboratory of Quantum Science and Engineering, Southern University of Science and Technology, Shenzhen 518055, China}
	
	\author{Ji Chu}
	\affiliation{Shenzhen Institute for Quantum Science and Engineering, Southern University of Science and Technology, Shenzhen 518055, China}
	\affiliation{International Quantum Academy, Shenzhen 518048, China}
	\affiliation{Guangdong Provincial Key Laboratory of Quantum Science and Engineering, Southern University of Science and Technology, Shenzhen 518055, China}
	
	\author{Jiawei Qiu}
	\affiliation{Shenzhen Institute for Quantum Science and Engineering, Southern University of Science and Technology, Shenzhen 518055, China}
	\affiliation{International Quantum Academy, Shenzhen 518048, China}
	\affiliation{Guangdong Provincial Key Laboratory of Quantum Science and Engineering, Southern University of Science and Technology, Shenzhen 518055, China}
	
	\author{Xuandong Sun}
	\affiliation{Shenzhen Institute for Quantum Science and Engineering, Southern University of Science and Technology, Shenzhen 518055, China}
	\affiliation{International Quantum Academy, Shenzhen 518048, China}
	\affiliation{Guangdong Provincial Key Laboratory of Quantum Science and Engineering, Southern University of Science and Technology, Shenzhen 518055, China}

	\author{Ziyu Tao}
	\affiliation{Shenzhen Institute for Quantum Science and Engineering, Southern University of Science and Technology, Shenzhen 518055, China}
	\affiliation{International Quantum Academy, Shenzhen 518048, China}
	\affiliation{Guangdong Provincial Key Laboratory of Quantum Science and Engineering, Southern University of Science and Technology, Shenzhen 518055, China}	
	
	\author{Jiawei Zhang}
	\affiliation{Shenzhen Institute for Quantum Science and Engineering, Southern University of Science and Technology, Shenzhen 518055, China}
	\affiliation{International Quantum Academy, Shenzhen 518048, China}
	\affiliation{Guangdong Provincial Key Laboratory of Quantum Science and Engineering, Southern University of Science and Technology, Shenzhen 518055, China}
	
	\author{Libo Zhang}
	\affiliation{Shenzhen Institute for Quantum Science and Engineering, Southern University of Science and Technology, Shenzhen 518055, China}
	\affiliation{International Quantum Academy, Shenzhen 518048, China}
	\affiliation{Guangdong Provincial Key Laboratory of Quantum Science and Engineering, Southern University of Science and Technology, Shenzhen 518055, China}

	\author{Yuxuan Zhou}
	\affiliation{Shenzhen Institute for Quantum Science and Engineering, Southern University of Science and Technology, Shenzhen 518055, China}
	\affiliation{International Quantum Academy, Shenzhen 518048, China}
	\affiliation{Guangdong Provincial Key Laboratory of Quantum Science and Engineering, Southern University of Science and Technology, Shenzhen 518055, China}
	
	\author{Yuanzhen Chen}
	\affiliation{Shenzhen Institute for Quantum Science and Engineering, Southern University of Science and Technology, Shenzhen 518055, China}
	\affiliation{International Quantum Academy, Shenzhen 518048, China}
	\affiliation{Guangdong Provincial Key Laboratory of Quantum Science and Engineering, Southern University of Science and Technology, Shenzhen 518055, China}
	\affiliation{Department of Physics, Southern University of Science and Technology, Shenzhen 518055, China}

	\author{Yang Liu}
	\affiliation{Shenzhen Institute for Quantum Science and Engineering, Southern University of Science and Technology, Shenzhen 518055, China}
	\affiliation{International Quantum Academy, Shenzhen 518048, China}
	\affiliation{Guangdong Provincial Key Laboratory of Quantum Science and Engineering, Southern University of Science and Technology, Shenzhen 518055, China}

	\author{Song Liu}
	\affiliation{Shenzhen Institute for Quantum Science and Engineering, Southern University of Science and Technology, Shenzhen 518055, China}
	\affiliation{International Quantum Academy, Shenzhen 518048, China}
	\affiliation{Guangdong Provincial Key Laboratory of Quantum Science and Engineering, Southern University of Science and Technology, Shenzhen 518055, China}

    \author{Youpeng Zhong}
    \email{zhongyp@sustech.edu.cn}
    \affiliation{Shenzhen Institute for Quantum Science and Engineering, Southern University of Science and Technology, Shenzhen 518055, China}
    \affiliation{International Quantum Academy, Shenzhen 518048, China}
    \affiliation{Guangdong Provincial Key Laboratory of Quantum Science and Engineering, Southern University of Science and Technology, Shenzhen 518055, China}
    
    \author{Jian-Jian Miao}
    \email{mjjphy@gmail.com}
    \affiliation{Department of Physics and HKU-UCAS Joint Institute for Theoretical and Computational Physics at Hong Kong, The University of Hong Kong, Hong Kong, China}

    \author{Jingjing Niu}
	\email{niujj@iqasz.cn}
	\affiliation{International Quantum Academy, Shenzhen 518048, China}

    \author{Dapeng Yu}
    \affiliation{Shenzhen Institute for Quantum Science and Engineering, Southern University of Science and Technology, Shenzhen 518055, China}
    \affiliation{International Quantum Academy, Shenzhen 518048, China}
    \affiliation{Guangdong Provincial Key Laboratory of Quantum Science and Engineering, Southern University of Science and Technology, Shenzhen 518055, China}
    \affiliation{Department of Physics, Southern University of Science and Technology, Shenzhen 518055, China}

\begin{abstract}
Fock-state lattices (FSLs), composed of photon number states with infinite Hilbert space, have emerged as a promising platform for simulating high-dimensional physics due to their potential to extend into arbitrarily high dimensions. 
Here, we demonstrate the construction of multi-dimensional FSLs using superconducting quantum circuits. By controlling artificial gauge fields within their internal structures, we investigate flux-induced extreme localization dynamics, such as Aharonov-Bohm caging, extending from 2D to 3D. We also explore the coherent interference of quantum superposition states, achieving
extreme localization within specific subspaces assisted by quantum entanglement. 
Our findings pave the way for manipulating the behavior of a broad class of quantum states in higher-dimensional systems. 
\end{abstract}
\maketitle

By reinterpreting degrees of freedom in quantum systems as additional spatial dimensions, synthetic dimensions have emerged as a powerful tool for exploring high-dimensional physics~\cite{Boada2012,Lohse2018,Zilberberg2018,Bouhiron2024,Lin2018}, offering new insights and experimental possibilities beyond the constraints of traditional spatial dimensions~\cite{Lustig2019, Ozawa2019,Dutt2020, Xiang2023,ArgueelloLuengo2024}.
A distinctive feature of synthetic dimensions is the ability to incorporate artificial gauge fields via site-dependent hopping~\cite{Celi2014,Tzuang2014,Fang2012}, extending phenomena traditionally limited to charged particles, such as the Aharonov-Bohm (AB) effect~\cite{Aharonov1959, Jaksch2003, Dalibard2011}, to neutral particles like photons~\cite{Tzuang2014, Lu2014, Ozawa2019a}. 
This ability facilitates the exploration of quantum transport and localization dynamics under gauge fields within synthetic lattices.
An intriguing phenomenon is AB caging, a flat-band localization effect~\cite{Vidal1998,Vidal2000,Vidal2001,Mosseri2022} arising from the combined influence of local topology~\cite{Sutherland1986} and magnetic fields.
Recent advancements in synthetic quantum systems have enabled the implementation of intricate lattice geometries~\cite{Koh2024,Yuan2018} and artificial gauge fields~\cite{Roushan2017,Li2023,Wang2024,Hung2021,Kanungo2022,Chen2024a}, making them ideal platforms for exploring AB caging in flat-band lattice. This phenomenon has been demonstrated in a variety of systems, including ion traps~\cite{Bermudez2011}, photonic lattices~\cite{Longhi2014,Mukherjee2018,Kremer2020}, circuit quantum electrodynamics~\cite{Martinez2023,Chen2024,Rosen2024np}, topolectrical circuits~\cite{Zhang2023}, ultracold atoms~\cite{Li2022}, and Rydberg lattices~\cite{Chen2024}, etc.  

Thus far, AB caging has been confined to two-dimensional (2D) systems, where wavefunctions are localized on finite subsets due to destructive interference, rather than exponential decay over the lattice sites. However, maintaining such extreme localization~\cite{Cherdantsev2018,Roentgen2018} becomes increasingly difficult as the coordination increases in higher dimensions, where more pathways allow the wavefunction to spread. In a 3D lattice, for example, particles initially localized on a 2D plane can escape to perpendicular space through available pathways outside the plane. Achieving extreme localization, or AB caging, in higher-dimensional systems not only challenges the precision of quantum state manipulation but also offers a platform for controlling single-particle and multi-particle dynamics in arbitrary-dimensional spaces.
Fock-state lattices (FSLs)~\cite{Deng2022,Saugmann2023,Yuan2024}, constructed from photon number states, offer a promising platform for simulating high-dimensional physics~\cite{Wang2016,Yao2023}. One of their key advantages is the ability to accommodate both a minimal zero-photon state and an unrestricted size range, enabling the construction of high-dimensional lattice symmetries with flexible energy-level structures. Additionally, by synthesizing artificial gauge fields within FSLs, these systems provide a unique platform for investigating phenomena such as  
exotic topological phases with specific boundary conditions~\cite{Cai2020,Mohanta2023}, as well as extreme localization in high-dimensional spaces.

Superconducting quantum circuits~\cite{Roushan2017,Ma2019,Wang2019,Yan2019,Gong2021,Saxberg2022,Morvan2022,Tao2023}, comprising linear oscillators and non-linear Josephson junctions, provide a versatile platform for constructing synthetic FSLs and exploring high-dimensional quantum phenomena~\cite{Wang2016, Cai2020, Zhang2022a, Yao2023}. 
In this work, we utilize superconducting qutrits to construct multi-dimensional synthetic FSLs, using 2D plaquettes as the building blocks. We demonstrate 2D AB caging on a single plaquette and extend this concept to 3D by incorporating two perpendicular rhombic plaquettes to form a 3D octahedral FSL. 
In this 3D system, localization arises from a distinct topological configuration formed by these two plaquettes with both half a flux quantum embedded in 3D space.
This extension is facilitated by the flexible energy-level configuration of qutrits, and we further show that the octahedral lattice can serve as a building block for constructing more complex FSLs involving additional qutrits.
Given the challenges of directly simulating real 3D space, our demonstration of 3D AB caging within a synthetic FSL confirms the feasibility of simulating quantum physics in higher synthetic dimensions.

\begin{figure}[t]
\begin{center}
    \includegraphics[width=0.45\textwidth]{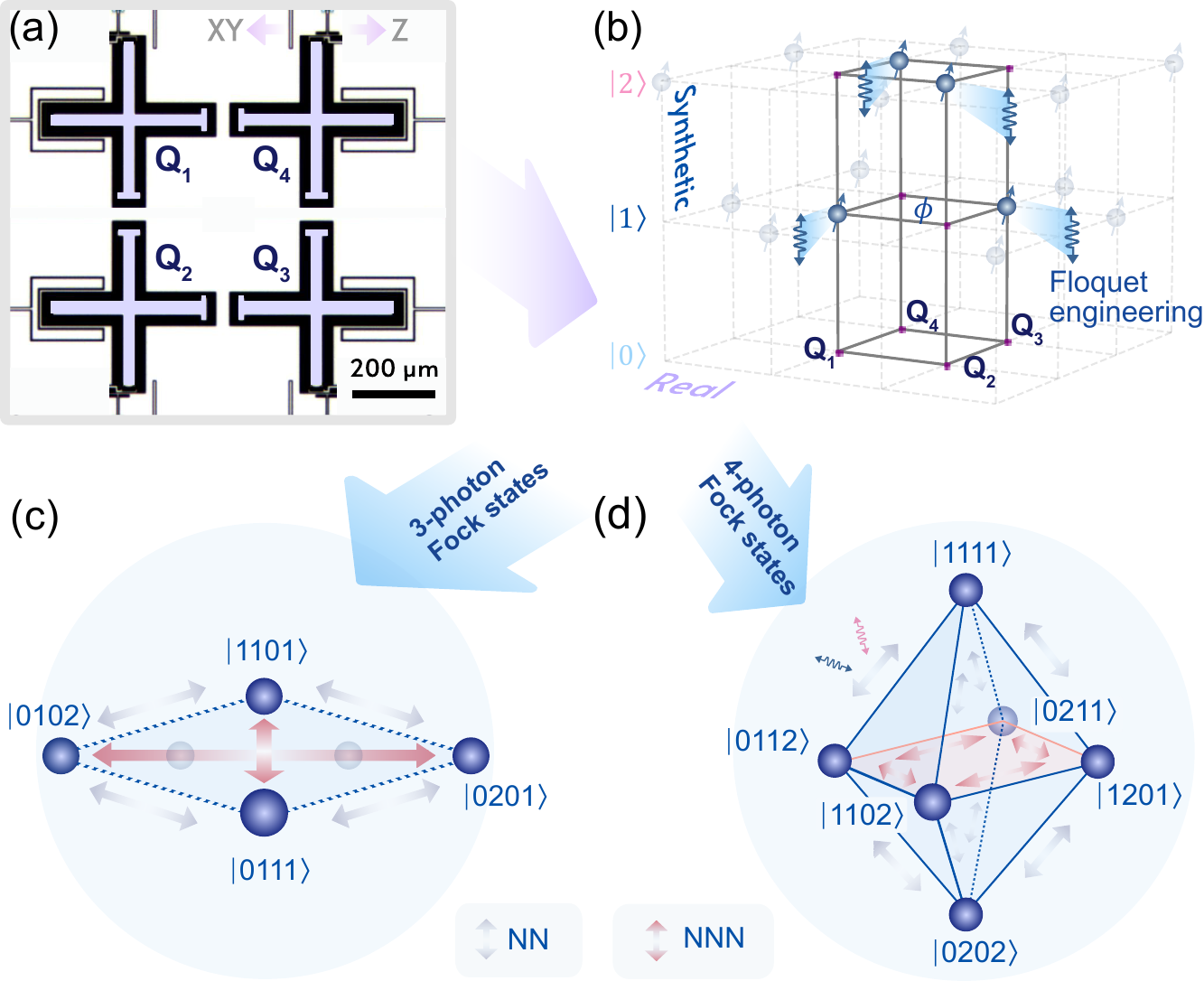}
    \caption{
    \label{fig1}
    {Synthetic AB cage in multi-photon FSLs.}
    (a) Photograph of the four-qutrit quantum processor. 
    (b) Schematic of the experiment.
    The $z$-axis represents the eigenstates in synthetic space, with spin symbols marking the excited states in real space. 
    The artificial gauge flux $\phi$ is synthesized via Floquet engineering of the qutrit frequencies. 
    (c) Schematic of a 2D synthetic plaquette constructed with three-photon Fock states.  
    (d) Schematic of a 3D synthetic octahedron constructed with four-photon Fock states.
    Blue arrows: NN coupling; red arrows: NNN coupling between qutrits.
    \vspace{-10pt}
    }
\end{center}
\end{figure}

The synthetic AB cages are implemented on a device containing four superconducting qutrits $Q_i$ ($i=1-4$) of the transmon type~\cite{Niu2023}, situated at corners of a plaquette and capacitively coupled to each other, as shown in \rfig{fig1}(a).
Extending the transmon Hilbert space to three levels allows us to engineer a broader range of synthetic lattice structures using fewer physical resources.
Qudits offer even more theoretical advantages, but they pose more stringent requirements on microwave control and readout.
The system's Hamiltonian is given by
\vspace{-5pt}
\begin{equation}\small\label{Hami}
    \frac{H}{\hbar}\!=\!\sum_{i=1}^4\!\left[\omega_{i}a_{i}^{\dagger}a_{i}\!+\!\frac{1}{2}U_{i}n_{i}\left(n_{i}\!-\!1\right)\right]\! +\!\sum_{i<j}g_{ij}(a_{i}^{\dagger}a_{j}\!+\!a_{j}^{\dagger}a_{i}),
\end{equation}
where $a_{i}$ and $n_{i}=a_{i}^{\dagger}a_{i}$ denote the bosonic annihilation and particle number operators. We truncate each operator at two photons and obtain a four-qutrit system.
For each qutrit $Q_i$, $\omega_{i}/2\pi$ is the resonant frequency between the lowest two eigenenergies, $U_{i}/2\pi\approx-216$~MHz represents the on-site attractive Hubbard interaction strength determined by the anharmonicity of the qutrit~\cite{Koch2007}, and $g_{ij}$ denotes the coupling strength between qutrits which are directly coupled through their capacitor pads, resulting in nearest-neighbor (NN) coupling strengths $g_\mathrm{12}/2\pi$, $g_\mathrm{23}/2\pi$, $g_\mathrm{34}/2\pi$, $g_\mathrm{41}/2\pi \sim14$ MHz, and next-nearest-neighbor (NNN) coupling strengths $g_\mathrm{13}/2\pi$, $g_\mathrm{24}/2\pi \sim3.2$ MHz. 
The multiphoton FSLs formed by the eigenstates of the qutrits facilitate the construction of multi-dimensional geometric configurations with only a few physical components. An interior artificial gauge field is synthesized by Floquet engineering~\cite{Shirley1965,Oka2019}, as illustrated in \rfig{fig1}(b).
This technique is implemented by applying periodic parametric driving to the qutrit on-site energy $\omega_i/2\pi$ through the Z control line.
Under this driving, the system's physics is captured by the following effective Hamiltonian within the FSL:
\begin{eqnarray}\small\label{Heff}
	H_{\mathrm{eff}}/\hbar=\sum_{k<l}\left(J_{kl}e^{i\phi_{kl}} |\psi_k\rangle \langle \psi_l|+H.c.\right),
\end{eqnarray}
where $J_{kl}$ denotes the effective hopping strength between states $|\psi_k\rangle$ and $|\psi_l\rangle$  at site $k$ and $l$ in synthetic space, tunable by the amplitude of the driving, and $\phi_{kl}$ is determined by the phase of the driving, see Ref.~\cite{SM2024} for details. 
These phases from parametric driving yield synthetic magnetic flux invariant under gauge transformation
$\Phi_{\mathcal{C}}=\sum_{\mathcal{C}}\phi_{kl}$ summed over the closed path of a Wilson loop $\mathcal{C}$~\cite{Wilson1974,Wang2020}.

We first illustrate the construction of a 2D plaquette using three-photon Fock states $|0111\rangle$, $|1101\rangle$, $|0102\rangle$, and $|0201\rangle$ as lattice sites $|\psi_k\rangle$ for $k=1$ to 4, where $Q_1$ and $Q_3$ operate within the $\{|0\rangle,|1\rangle \}$ subspace, while $Q_2$ and $Q_4$ operate within the $\{|1\rangle,|2\rangle \}$ subspace, as illustrated in \rfig{fig1}(c) (see details in Ref.~\cite{SM2024}). 
Specifically, the qutrit frequencies $\omega_2/2\pi$ and $\omega_4/2\pi$, are quenched to approximately $4.91$~GHz, while $\omega_1/2\pi$ and $\omega_3/2\pi$ are tuned to approximately $4.73$ GHz, where the energy levels of $|11\rangle$ and $|02\rangle$ of each pair of NN qutrits are quenched to the vicinity of the operational frequencies~\cite{SM2024}, then the qutrit frequencies are modulated by parametric driving to establish the synthetic lattice. The nonlinearity $U_i$ of our superconducting qutrits prevents state leakage outside this synthetic subspace. We can further extend this approach to explore higher-dimensional geometries, constructing a 3D lattice with four-photon Fock states:  $|0202\rangle$, $|0112\rangle$, $|1201\rangle$, $|0211\rangle$, $|1102\rangle$, and $|1111\rangle$, forming a octahedral structure (\rfig{fig1}(d)). Upon tuning the flux in the synthetic space via Floquet engineering, the states traversing different paths accumulate distinct phases, leading to constructive or destructive interference phenomena.

\begin{figure}[t]
\begin{center}
	\includegraphics[width=0.45\textwidth]{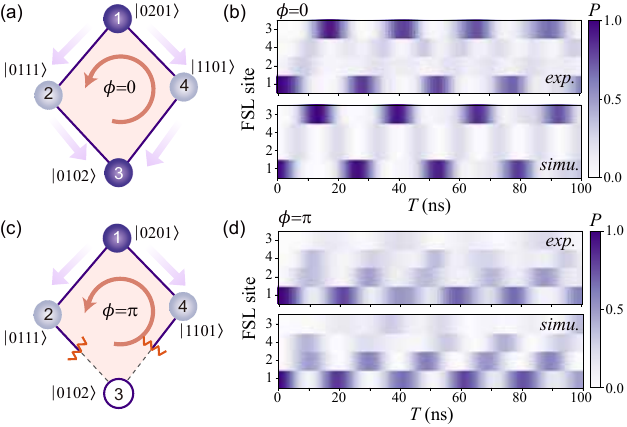}
	\caption{
    \label{fig2}
    (a) Diagram of a free quantum walk in a closed loop.
    (b) Constructive interference at $\phi=0$, resulting in identical hopping rates through two paths. 
    (c) Illustration of the AB cage phenomenon in a synthetic 2D plaquette.
    (d) Destructive interference at $\phi=\pi$, with hopping through both paths retained.
    The top and bottom panels in (b) and (d) display experimental data and numerical simulations, respectively.
    \vspace{-10pt}
    }
\end{center}
\end{figure}

Now we present the experimental demonstration of AB caging in a 2D synthetic plaquette.
The system is initialized in the three-photon Fock state $|\psi_1\rangle=|0201\rangle$ on a plaquette (\rfig{fig2}(a)). The qutrits are then quenched to their operational frequencies to establish the synthetic space (Ref.~\cite{SM2024} Sec. III). 
Subsequently, we apply parametric driving on all qutrits to form a dual-path interferometer.
The NN hopping strengths ($J_{12}/2\pi$, $J_{23}/2\pi$, $J_{34}/2\pi$, $J_{41}/2\pi$) are tuned to $\sim18.4$ MHz, and phase between lattice sites are adjusted by tuning the driving parameters on the two qutrits involved in the interaction between the $|11\rangle$ and $|02\rangle$ states (see details in Ref.~\cite{SM2024}).
When the synthetic flux $\phi=\phi_{12}+\phi_{23}+\phi_{34}+\phi_{41}$ is tuned to zero (\rfig{fig2}(a)), i.e., by setting $\phi_{12}=\phi_{23}=\phi_{34}=\phi_{41}=0$, and the transition strengths along the four paths are balanced, constructive interference results in a free quantum walk between the state $|0201\rangle$ at site 1 and state $|\psi_3\rangle=|0102\rangle$ at site 3, see the evolution of the site population $P$ versus the evolution time $T$ in \rfig{fig2}(b). 
Conversely, when the synthetic flux is tuned to $\phi=\pi$ by adjusting $\phi_{12}=\phi_{23}=\pi/2$ and $\phi_{34}=\phi_{41}=0$, destructive interference erases population at site 3 (see \rfigs{fig2}(c) and (d)).
This extreme localization pattern confirms the establishment of an AB cage within 2D plaquette. 
The observed residual population at site 3 is attributed to experimental imperfections in the actual driving Hamiltonian (see \rfig{fig2}{(d)} and Ref.\cite{SM2024}, Sec. III.A.)

\begin{figure*}[t]
\begin{center}
	\includegraphics[width=0.9\textwidth]{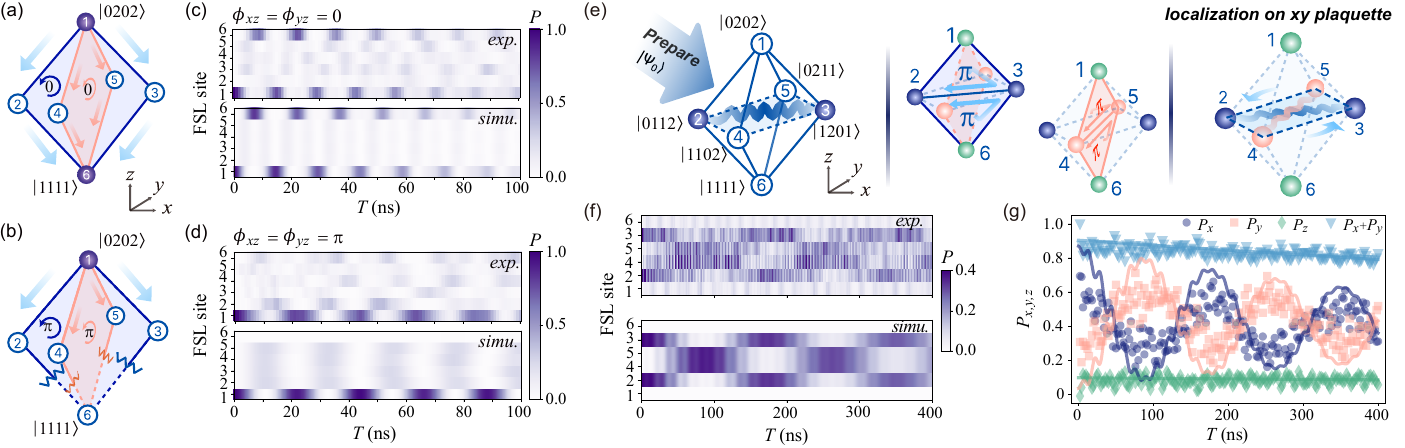}
	\caption{
    \label{fig3new}
    (a) Schematic of two plaquettes with a common $z$-axis, the sites $1-6$ represent the states $|0202\rangle$, $|0112\rangle$, $|1201\rangle$, $|1102\rangle$, $|0211\rangle$, and $|1111\rangle$, respectively.
    (b) Pseudo-3D AB cage with flux in both plaquettes set to $\pi$.
    (c) Evolution dynamics among the six sites, akin to the 2D plaquette. With flux in both plaquettes set to zero, a free quantum walk through four paths occurs, resulting in state transfer between $|0202\rangle$ and $|1111\rangle$.
    (d) Destructive interference along the $z$-axis, leading to population cancellation at $|1111\rangle$.
    (e) Schematic of 3D AB cage dynamics on a skewed octahedral FSL. Left: Initialization to a two-site superposition state, $|\Psi_4\rangle = \frac{1}{\sqrt{2}}(|0112\rangle+|1201\rangle)$. Middle: Two vertical plaquettes each trapping a $\pi$ flux within two triangular regions, resulting in destructive interference for both plaquettes. Right: Photon evolution is confined to the equatorial section of the 3D AB cage.
    (f) Evolution dynamics among the six sites in the 3D AB cage. 
    The top and bottom panels in (c), (d), and (f) show experimental data and numerical simulations, respectively.
    (g) Evolution of probabilities $P_{x,y,z}$, illustrating the localization pattern. 
    Dots: experimental results; solid curves: numerical simulations. 
    \vspace{-10pt}
    }
\end{center}
\end{figure*}

Next, we extend our synthetic framework beyond 2D. In real space, expanding the dynamics of AB cages typically requires increasing the number of atoms and altering their arrangement. However, in our synthetic FSL, this can be achieved by lifting the number of photons constituting the FSL. Specifically, 
 we manipulate six Fock states of four-photon excitations in resonance: $|0202\rangle$, $|1201\rangle$, $|1102\rangle$, $|0211\rangle$, $|0112\rangle$, and $|1111\rangle$, sequentially labeled from synthetic sites 1 to 6, as depicted in \rfigs{fig3new}(a) and (c). 
In this configuration, sites 1, 2, 3, and 6 form one plaquette, while sites 1, 4, 5, and 6 form another, encompassing both left-right and front-back paths. 
These two plaquettes share a common $z$-axis, resulting in a composite interference pattern that includes both the $xz$ (blue) and $yz$ (red) plaquettes. 
The synthetic flux exists in both plaquettes, jointly controlling the state transport behavior along the paths formed by the NN $|11\rangle - |02\rangle$ coupling. The fluxes are $\phi_{xz}=\phi_{12}+\phi_{26}+\phi_{63}+\phi_{31}$ and $\phi_{yz}=\phi_{14}+\phi_{46}+\phi_{65}+\phi_{51}$.
The system is initialized at site 1 ($|\psi_1\rangle=|0202\rangle$) with qutrits at idling frequencies. Frequency modulation, similar to the 2D AB cage experiment, establishes synthetic phases across the two plaquettes. 
When the flux within both plaquettes is set to zero, namely $\phi_{xz}=\phi_{yz}=0$, and the hopping strengths along the four paths are balanced, we observe a free quantum walk between site 1 and site 6 ($|1111\rangle$), as demonstrated in \rfig{fig3new}(b).
Conversely, setting $\phi_{xz}=\phi_{yz}=\pi$, by tuning $\phi_{12}=\phi_{63}=\phi_{46}=\phi_{51}=\pi/2$, and $\phi_{26}=\phi_{31}=\phi_{14}=\phi_{65}=0$, induces destructive interference across all four paths, leading to localization on site 1 (\rfig{fig3new}(d)).
Experimental results align well with numerical simulations, the slight distortions appear at sites 2, 3, 4, and 5 could be caused by imbalanced hopping strengths across all four paths during parametric modulation.
The desynchronization of probability distribution along different paths and slight delocalization on site 6 are likely caused by both the disorder in hopping strength and phase mismatch due to shifts in energy levels when modulating the phase of parametric driving pulses (see~\cite{SM2024} Fig.S10 for numerical simulation analysis). 
To mitigate these effects, we have slightly decreased the parametric drive amplitude in this scenario.

The extension towards pseudo-3D demonstrates that synthetic magnetic fields can still be simultaneously controlled within an expanded synthetic space. However, this verification is limited to the combined effects of interference paths, similar to the 2D scenario. Our next step will utilize quantum coherence and superposition to explore the richer dynamics within a genuine 3D cage. 
By activating NNN coupling, we extend the dual-plaquette structure into a complete 3D octahedral configuration, as illustrated in \rfig{fig3new}(e). 
Careful tuning of the resonance positions of synthetic sites 2, 3, 4, and 5 establishes NNN hopping among these sites, forming a plaquette perpendicular to the $z$-axis, termed the $xy$ plaquette. 
This equatorial plaquette comprises sites 2, 3, 4, and 5 (left in \rfig{fig3new}(e)), with NNN hopping strengths satisfying $J_{24}=J_{35}=J_{25}/2=J_{34}/2$, where $J_{34}/2\pi \sim 5.7$ MHz, making this octahedron ``skewed'', see~\cite{SM2024} for details.
Previously, we have discussed the dynamics in 2D plaquettes with the system initialized at a single site, showing the features of a multi-path interferometer but only resulting in dynamics and localization at a single site. A full exhibition of 3D AB caging requires the capability of localization on a plane. To manifest this feature, we initialize the system in a quantum entangled state within the synthetic 3D FSL to directly control the superposition state interference and explore caging dynamics distinguished from the 2D cases.

\begin{figure*}[hbt]
\begin{center}
    \includegraphics[width=0.9\textwidth]{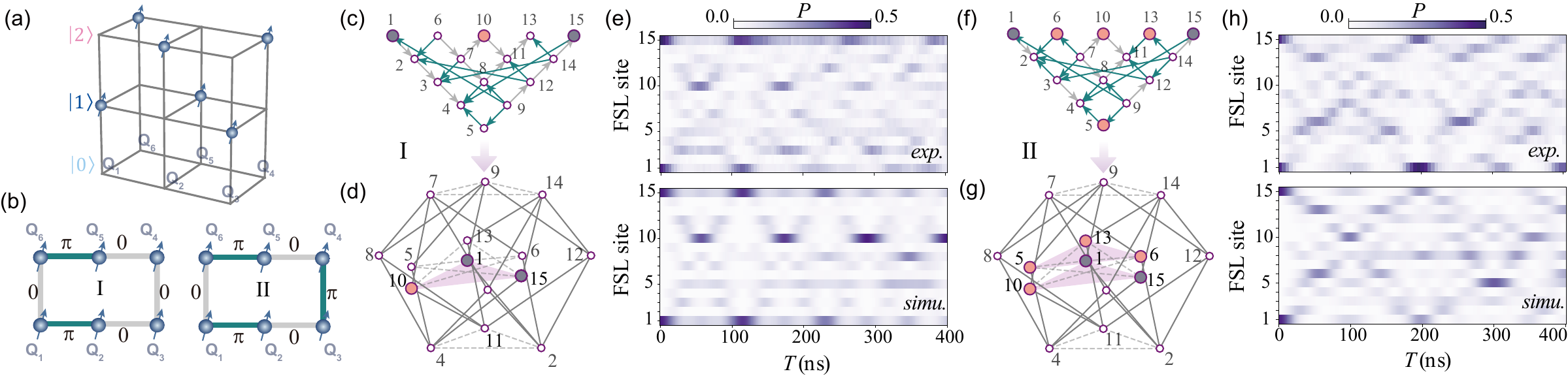}
	\caption{
    \label{fig4new}
    (a) Energy level configuration of the 6-qutrit loop. 
    (b) Flux modulation configurations (I and II) on the loop edges.
    (c-d) Synthetic FSL formed in the two-excitation subspace with flux configuration I. Solid lines: NN coupling; dashed lines: NNN coupling. Blue circles: initial sites; pink circles: sites with positive superposition in different interference paths. (c) synthetic sites and NN coupling, (d) effective FSL folded by the actual coupling strengths.
    (e) Dynamics of superposition state $|\Psi_6\rangle=(|120101\rangle+|010112\rangle)/\sqrt{2}$ in the FSL from (c). 
    (f-g) Synthetic FSL with flux configuration II.
    (h) Dynamics of $|\Psi_6\rangle$ in FSL from (f). 
    Upper and lower panels in (e) and (h) show experimental data and numerical simulations, respectively.
    \vspace{-10pt}
    }
\end{center}
\end{figure*}

To experimentally realize this, we initialize the system in the superposition state at sites 2 and 3: $|\psi_0\rangle = \frac{1}{\sqrt{2}} (|0112\rangle + |1201\rangle)$, as shown in \rfig{fig3new}(e).
Subsequently, we apply parametric driving to the qutrits similar to the previous experiments. 
We first focus on the $xz$ plane (comprising sites 1, 2, 3, and 6). By setting $\phi_{12}=\phi_{63}=\pi$ and $\phi_{26}=\phi_{31}=0$, the plaquette is divided into two triangular regions, each containing a $\pi$ synthetic flux (middle in \rfig{fig3new}(e)). As a consequence, the superposition state becomes confined at sites 2 and 3.
Applying the same phase modulation process to the $yz$ plaquette confines the system to the superposition state at sites 4 and 5, and the combined modulations on $xz$ and $yz$ plaquettes freeze the dynamical evolution along the $z$ axis. Within these operations, we adjust the operational frequencies of four qutrits to keep sites 1--6 in resonance, therefore no extra flux is trapped inside the $xy$ plaquette. The NNN coupling induces constructive interference at sites 2, 3, 4, and 5, as shown in the lower panel of \rfig{fig3new}(f), the synchronized dynamics between sites 2, 3 and sites 4, 5 with evolution time $T$, indicates that quantum superposition is coherently maintained in this 3D AB cage. This coherent localization behavior can also be understood by the FSL site population projected on three directions, where the $x$-component, $P_x = P_2 + P_3$, and $y$-component, $P_y = P_4 + P_5$, undergo typical vacuum Rabi oscillations, here $P_k$ means the population of state $|\psi_k\rangle$ at site $k$. The combined quantity $P_x + P_y$ characterizes the distribution of localized states in the $xy$ plaquette (right in \rfig{fig3new}(g)), while the $z$-component ($P_z = P_1 + P_6$) remains low occupation, indicating that the residual distribution outside the $xy$   plaquette is extremely caged.
These experimental results demonstrate the successful synthesis of magnetic flux across two longitudinal sections ($xz$ and $yz$ plaquettes) within a 3D AB cage, which effectively localizes a quantum superposition state within the $xy$ plaquette. This effect provides evidence for the realization of a genuine 3D AB cage with the assistance of NNN coupling.

Our method demonstrates a promising feature that more complex FSLs can be synthesized with only a few physical resources. As an example, we extend our investigation to six qutrits on another quantum processor (see Ref.~\cite{SM2024} Fig.S14). The operational frequencies of the NN qutrits are staggered (see \rfig{fig4new}{(a)}), in the same approach as the previous experiment, facilitating the formation of hopping terms through $11-02$ coupling. The phase of each coupling between two qutrits is independently tunable. The system's ground state is $|010101\rangle$, with higher-frequency qutrits initialized to $|1\rangle$.
Based on this ground state, the FSL consists of 15 two-excitation Fock states, labeled as sites 1 to 15: $|120101\rangle$, $|111101\rangle$, $|110201\rangle$, $|110111\rangle$, $|110102\rangle$, $|021101\rangle$, $|020201\rangle$, $|020111\rangle$, $|020102\rangle$, $|011201\rangle$, $|011111\rangle$, $|011102\rangle$, $|010211\rangle$, $|010202\rangle$, and $|010112\rangle$ (see Ref.~\cite{SM2024} Section IV.A).
The dynamics of these states are governed by coherent interference modulated by the applied flux configurations. 
We illustrate two specific flux configurations, I and II, in \rfig{fig4new}{(b)}, with corresponding state distributions in \rfig{fig4new}{(c)} and (f). 
The system begins in a superposition of sites 1 and 15, marked by solid blue circles. Arrows in \rfig{fig4new}{(c)} and (f) indicate the phase direction of multi-path interference (see Ref.~\cite{SM2024} Fig. S15). The \rfigs{fig4new}{(d)} and (g) show 3D projections of the FSL, where edge lengths correspond to the actual coupling strengths.
In flux configuration I, multi-path interference leads to a periodic positive superposition at site 10 (\rfig{fig4new}{(e)}), while populations at most other sites are canceled by destructive interference. The dynamics are thus localized in a ``plane'' formed by sites 1, 10 and 15 (marked in pink in \rfig{fig4new}{(d)}). In flux configuration II, periodic superpositions occur at sites 5, 6, 10, and 13 (\rfig{fig4new}{(h)}), forming a localized subspace consisting of two ``planes'' (marked in pink in \rfig{fig4new}{(g)}). 
These patterns can be understood as arising from multiple paths connecting the entangled states at sites 1 and 15, resulting in localization within subspaces of FSL sites with NN couplings (see Ref.~\cite{SM2024}, Section IV.B).
Additionally, we examine the evolution and interference when the system is initialized at single FSL sites, 1 or 15 (see Section IV.C of Ref.~\cite{SM2024}).  These results further demonstrate the ability of our approach to construct subspaces of compact localized states in more complex FSLs.

In conclusion, our work demonstrates the construction of multi-dimensional FSLs using superconducting qutrits. 
Utilizing Floquet engineering and tunable coupler, we effectively manipulate coupling strengths and create synthetic gauge fields in multi-photon FSLs, allowing us to explore gauge field-induced extreme localization dynamics such as AB caging from 2D to 3D. 
We also investigate the coherent interference of quantum superposition states, achieving extreme localization within specific subspaces of multi-dimensional AB cages, facilitated by quantum entanglement.
our work opens a pathway to localize a large family of states on FSLs under gauge fields in higher-dimensional systems, leveraging the internal degrees of freedom inherent in quantum platforms. This approach could be useful for studying quantum many-body systems, such as perovskite materials with high-temperature superconductivity~\cite{Bednorz1988,Maeno1994} and giant magnetoresistance~\cite{Moritomo1996}, topological edge states~\cite{Schindler2018,Xie2021rev}, and quantum chaos~\cite{Suntajs2020}.

\begin{acknowledgments}
\vspace{-10pt}
This work was supported by the National Natural Science Foundation of China (12174178, 12204228, 12374474), the Guangdong Provincial Key Laboratory (2019B121203002), the Science, Technology and Innovation Commission of Shenzhen Municipality (KQTD20210811090049034, K21547502), the Shenzhen-Hong Kong Cooperation Zone for Technology and Innovation (HZQB-KCZYB-2020050), and the Guangdong Basic and Applied Basic Research Foundation (2022A1515110615, 2024A1515011714).
J. J. Miao was supported by General Research Fund Grant No. 14302021 from Research Grants Council of Hong Kong.
\end{acknowledgments}


%

\end{document}


\title{Supplemental Material for ``Synthetic multi-dimensional Aharonov-Bohm cages in Fock state lattices''}

\maketitle

\setcounter{equation}{0}
\setcounter{figure}{0}
\setcounter{table}{0}
\setcounter{page}{1}

\tableofcontents
\newpage

\section{Experimental setup} 

\begin{figure}[ht]
\begin{center}
    \includegraphics[width=0.8\textwidth]{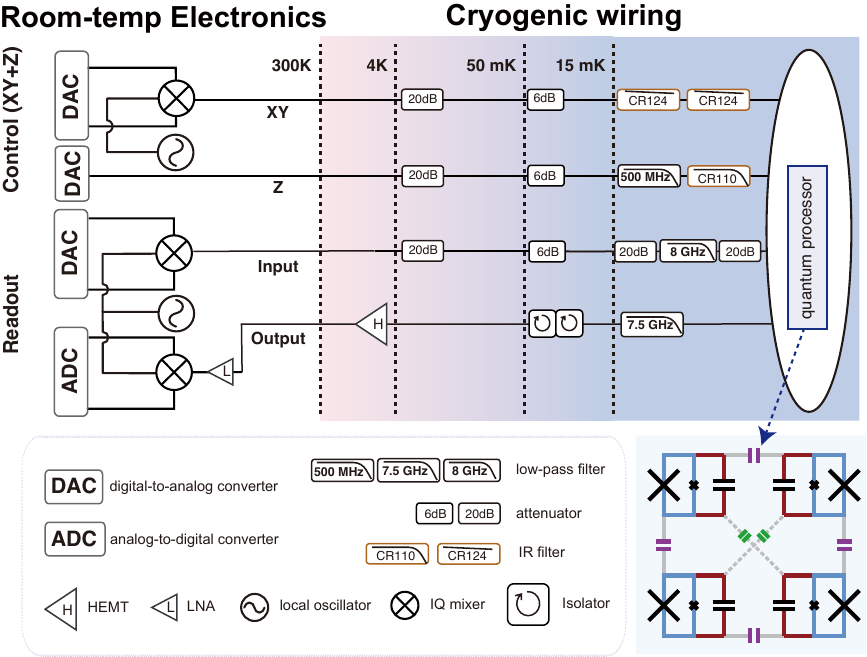}
    \caption{
        \label{sample}
        The experimental setup includes measurement circuit diagrams ranging from room temperature to low temperature. Bottom left: legend of the electronic components. Bottom right: circuit diagram of the quantum processor. The two Josephson junctions constituting the superconducting quantum interference device (SQUID) of the qutrits are asymmetric.
    }
\end{center}
\end{figure}
This work is implemented on the experimental setup as \rfig{sample}. The whole system includes the room temperature measurement and control electronics, a superconducting quantum processor inside a dilution refrigerator, and coaxial cable wiring extending from the electronics to the processor. The electronics features two types of modules: digital-to-analog (DAC) module and analog-to-digital (ADC) module, both are custom built for qubit control and readout~\cite{Tao2023}. The DAC module generates Z bias signals ranging from DC bias to 300 MHz intermediate frequency (IF) pulses or perform frequency up-conversion with a local oscillator and an IQ mixer for XY drive matching the qutrit resonance frequency. Conversely, the ADC module samples time-domain analog voltage signals transmitted from the readout line in the quantum processor, where the GHz microwave output signals are first down-converted to IF frequencies through another IQ mixing process, then digitized by the ADC and demodulated into a data point in the complex phase space. This sampling process is repeated at a programmable rate to reconstruct the probability distribution of the qutrit state.

\begin{table}[hbt]
    \centering
    \caption{Device information}
    \label{qutrit_info}
    \begin{tabular}{|p{4cm}<{\centering}|p{1.5cm}<{\centering}|p{1.5cm}<{\centering}|p{1.5cm}<{\centering}|p{1.5cm}<{\centering}|}
        \hline
	\hline	        
        \text{Qutrit} & $Q_1$ & $Q_2$  & $Q_3$  & $Q_4$  \\ \hline
        \text{$\omega_\mathrm{idle}/2\pi$ (GHz)} & 4.436 & 5.077 & 4.339 & 4.924 \\ \hline
        \text{$\omega_\mathrm{oper}/2\pi$ (GHz)} & 4.727 & 4.907 & 4.736 & 4.910 \\ \hline
        \text{$\omega_r/2\pi$ (GHz)} & 5.515 & 5.563 & 5.621 & 5.686 \\ \hline
        \text{$U_i/2\pi$ (MHz)} & -221.7 & -207.0 & -224.5 & -210.0 \\ \hline
        \text{$F_{00}$} & 0.908 & 0.95 & 0.944 & 0.763 \\ \hline
        \text{$F_{11}$} & 0.862 & 0.871 & 0.856 & 0.774\\ \hline
        \text{$F_{22}$} & 0.826 & 0.795 & 0.846 & 0.774\\ \hline
        \text{$T_1^{10} (\mu s)$} & 36.3 & 13.6 & 19.2 & 18.4\\ \hline
        \text{$T_2^{10}\ (\mu s)$} & 5.91 & 2.71 & 1.52 & 4.77\\ \hline
        \textbf{$T_1^{21}\ (\mu s)$} & 15.1 & 7.0 & 13.0 & 9.4\\ \hline
        \textbf{$T_2^{21}\ (\mu s)$} & 2.15 & 0.91 & 2.66 & 2.77\\ \hline
    \end{tabular}
\end{table}

\rtab{qutrit_info} presents the parameters and typical performance of the quantum processor. The four superconducting artificial atoms on the processor, used as``qutrits" $Q_1$--$Q_4$, are directly coupled through their capacitor pads, resulting in nearest-neighbor (NN) coupling strengths $g_\mathrm{12}/2\pi,g_\mathrm{23}/2\pi,g_\mathrm{34}/2\pi,g_\mathrm{41}/2\pi \sim 14$ MHz and next-nearest-neighbor (NNN) coupling strengths $g_\mathrm{13}/2\pi,g_\mathrm{24}/2\pi \sim3.2$ MHz. 
$\omega_\mathrm{idle}/2\pi$ denotes the idling frequency spacing between energy levels $|0\rangle$ and $|1\rangle$, $\omega_\mathrm{oper}/2\pi$ denotes the operating frequency of each qutrit to build the 2D plaquette.
$\omega_r/2\pi$ represents the driving and demodulation frequency for the readout resonator, selected to optimize the signal-to-noise ratio for distinguishing $|1\rangle$ and $|2\rangle$. $U_i/2\pi$ is the anharmonicity of $Q_i$, which is experimentally measured through Rabi driving between $|1\rangle$ and $|2\rangle$. $F_{00}$, $F_{11}$, and $F_{22}$ represent the state preparation and measurement (SPAM) fidelity of the three computational states ($|0\rangle$, $|1\rangle$ and $|2\rangle$).

\section{Experimental calibration}

\subsection{Fock state preparation and measurement}
\label{Fockstate_SPAM}
\begin{figure}[b]
\begin{center}
    \includegraphics[width=0.8\textwidth]{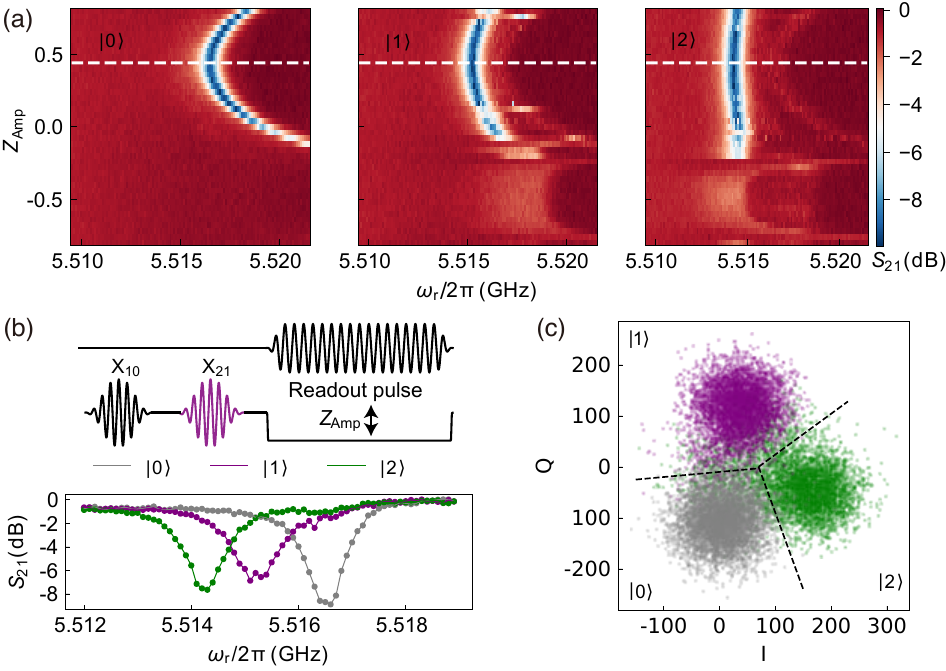}
    \caption{
        \label{Dispersive_readout}
     Dispersive readout of states $|0\rangle$, $|1\rangle$, and $|2\rangle$.
    (a) Transmission scan of the readout Z pulse amplitude ($Z_\mathrm{Amp}$) and readout frequency ($\omega_r$), extracting the transmission after preparing the qutrit separately in states $|0\rangle$, $|1\rangle$, and $|2\rangle$. White dashed lines indicate the chosen readout Z pulse amplitude for optimal visibility of all states.
    (b) Dispersive shift at the optimized readout parameters. Circuit diagram depicts the pulse sequence used in the measurement calibration experiment for $|2\rangle$.
    (c) Single-shot IQ measurement at the optimized readout parameters.
    }
\end{center}
\end{figure}
\begin{figure}[ht]
\begin{center}
    \includegraphics[width=0.65\textwidth]{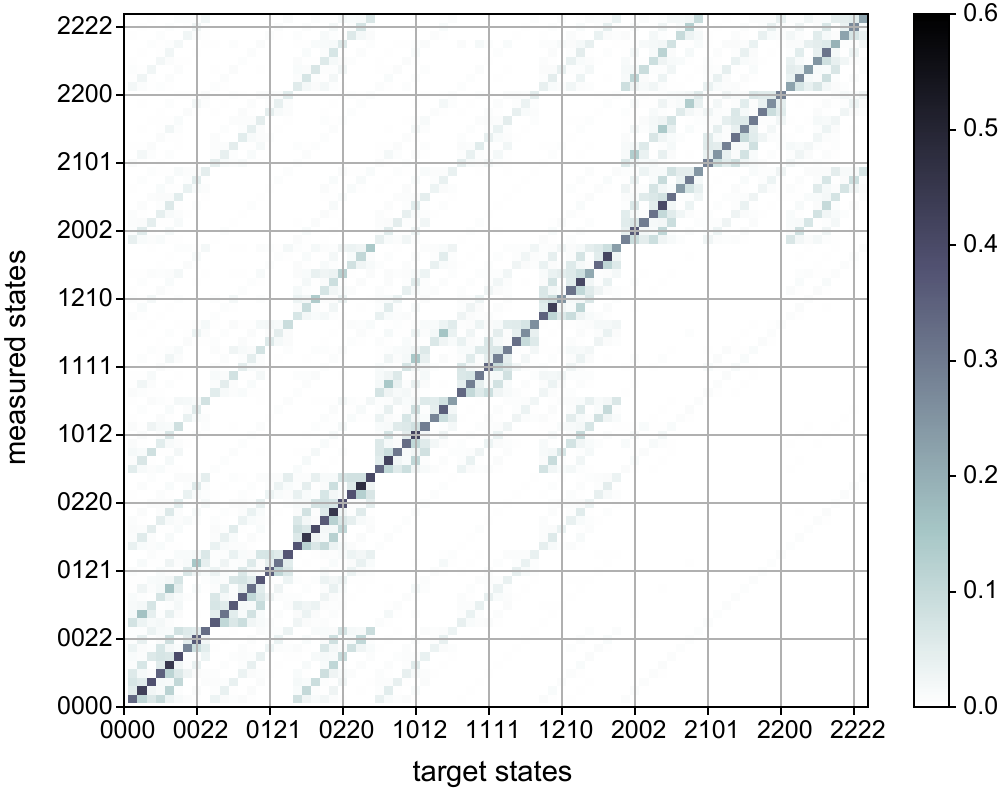}
    \caption{
        \label{ReadCrossMat}
        {The readout crosstalk matrix.}
        {The four-qutrit system is prepared and measured in the computational basis spanning 81 states ($i\otimes j \otimes m \otimes n$, with $i,j,m,n \in\{|0\rangle, |1\rangle,|2\rangle\}$). The colorbar indicates the state-preparation fidelity, while off-diagonal values represent crosstalk errors.}
    }
\end{center}
\end{figure}

Building the synthetic Fock state lattice (FSL) and measuring the caging dynamics require individual control and clear visibility of the multi-photon Fock states. However, the residual coupling strength among the multiphoton Fock states leads to unexpected population scattering onto non-targeted states during state preparation and measurement. To address this issue, we implemented two strategies. Firstly, we carefully tuned $\omega_{\mathrm{idle}}/2\pi$ to stagger the excited levels, ensuring that populations could be efficiently pumped onto all positions within the expected synthetic space.  Secondly, we adjusted the Z pulse amplitude applied simultaneously with the readout pulse to ensure reliable visibility of each expected Fock state. As illustrated in \rfig{Dispersive_readout}, we optimized the readout quality of each qutrit and achieve a clear visibility for each qutrit state. Furthermore, in \rfig{ReadCrossMat}, we present the measured probabilities of preparing four-qutrit bases: $i\otimes j \otimes m \otimes n$, with $i,j,m,n \in\{|0\rangle, |1\rangle,|2\rangle\}$ ($i.e.$, states $|0000\rangle$, $|0001\rangle$, ..., and $|2222\rangle$). This SPAM matrix was obtained after each experimental sequence, enabling an inverse transformation on the experimentally measured probabilities to correct for SPAM errors.

\subsection{Floquet engineering}
\label{paradrive}
Building synthetic multi-dimensional AB cages requires Floquet engineering~\cite{Shirley1965,Oka2019} to control hoppings between Fock states and create artificial gauge fields. The coupling between two capacitive-coupled superconducting qutrits is tunable through parametric driving on qutrit frequencies. When a time-periodic driving on the qutrit frequency is applied as $\omega(t) = \omega_{oper} + \Omega_d \cos(\omega_p t + \varphi)$, where $\omega_{oper}$ is the operational frequency, also act as the constant part of the qutrit frequency. $\omega_p$ denotes the driving frequency, $\Omega_d$ denotes the driving amplitude, $\varphi$ is the phase of parametric driving on this qutrit. In the rotating frame of two qutrits $Q_1$ and $Q_2$, the periodic driving induces an effective interaction Hamiltonian~\cite{Reagor2018a}: 
\begin{align}\small
H^{\mathrm{eff}}_{1,2} &= \sum_{n=-\infty}^{\infty} g_{12} J_n(\Omega_{d1}/\omega_p) e^{i(\varphi_1-\varphi_2)} \times \{e^{i(n\omega_p-\Delta_{12})t}|10\rangle \langle01| \nonumber \\ 
&+ \sqrt{2} [e^{i(n\omega_p-\Delta_{12}-U_2)t} |20\rangle \langle11| 
+ e^{i(n\omega_p-\Delta_{12}+U_1)t} |11\rangle \langle02|]\}  + h.c. \nonumber
\end{align}
Here, $\Delta_{12}=\bar{\omega}_{Q1}-\bar{\omega}_{Q2}$, where $\bar{\omega}_{Q}$ denotes the time-domain average of $\omega(t)$, $\varphi_1$ and $\varphi_2$ are the phases of driving on $Q_1$ and $Q_2$, respectively. $U_1$ and $U_2$ are the anharmonicities of two qutrits. $J_n$ represents the $n$-th first-kind Bessel function. $g_{12}$ is the actual capacitive coupling strength between $Q_1$ and $Q_2$. 
\begin{figure}[b]
\begin{center}
    \includegraphics[width=0.6\textwidth]{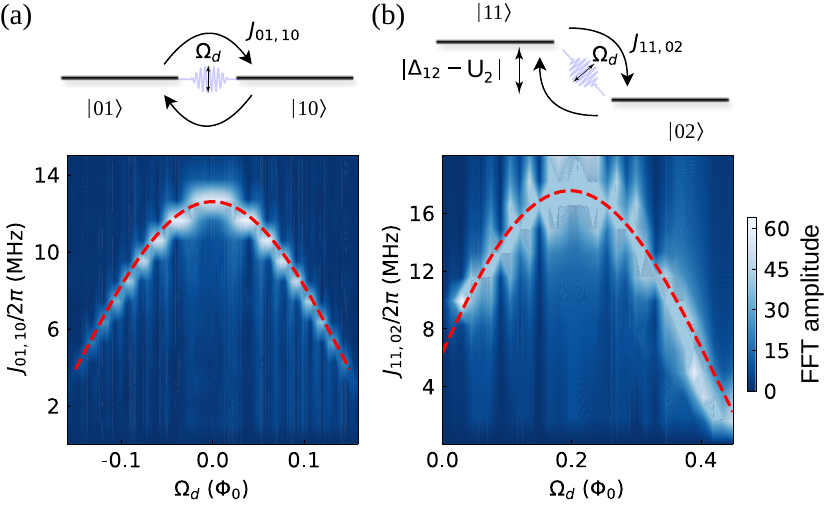}
    \caption{
        \label{paraamp_modu}
        {Modulating coupling strength with driving amplitude $\Omega_d$ (using quantum flux $\Phi_0$ as the unit).}
        {(a) Modulation of coupling between $|01\rangle$ and $|10\rangle$ ($J_{01,10}$) when two qutrits are in resonance.}
        {(b) Modulation of coupling between $|11\rangle$ and $|02\rangle$ ($J_{11,02}$) when the two qutrits are detuned by $\Delta_{12}/2\pi\sim 180 ~\mathrm{MHz}$.}
    }
\end{center}
\end{figure}

When $Q_1$ and $Q_2$ are resonant, the hopping between $|01\rangle$ and $|10\rangle$ $J_{\text{01,10}}\propto J_n(\Omega_d/\omega_p)$ can be modulated as shown in \rfig{paraamp_modu}(a). The white regions illustrate the hopping strength extracted through Fast Fourier Transformation (FFT) of the dynamical evolution, the red dashed curves are the fitting results. 
When $Q_1$ and $Q_2$ are detuned, considering the $\{|11\rangle,|02\rangle \}$ subspace, the coupling between $|11\rangle$ and $|02\rangle$ can be modulated as shown in \rfig{paraamp_modu}(b). The phase associated with this NN coupling can be approximately expressed as $\delta \varphi_{12}(t) \approx \varphi_1 - \varphi_2 + (n\omega_p - \Delta_{12} + U_1)t$. 
In this work, the NN hopping between two multi-photon Fock states is tuned through driving the two NN qutrits involved in their $|11\rangle-|02\rangle$ interaction. 
Specifically, for the 2D AB caging experiment, $Q_2$ and $Q_4$ are tuned to approximately $4.91$ GHz, $Q_1$ and $Q_3$ are tuned to approximately $4.73$ GHz, where NN $|11\rangle$ and $|02\rangle$ are detuned. For the pseudo-3D and 3D AB caging experiments, we shift $Q_2$ to approximately $\sim 4.88$ GHz, $Q_4$ to $\sim4.94$ GHz, $Q_1$ and $Q_3$ to $\sim 4.7$ GHz.

\begin{figure}[h]
\begin{center}
    \includegraphics[width=0.8\textwidth]{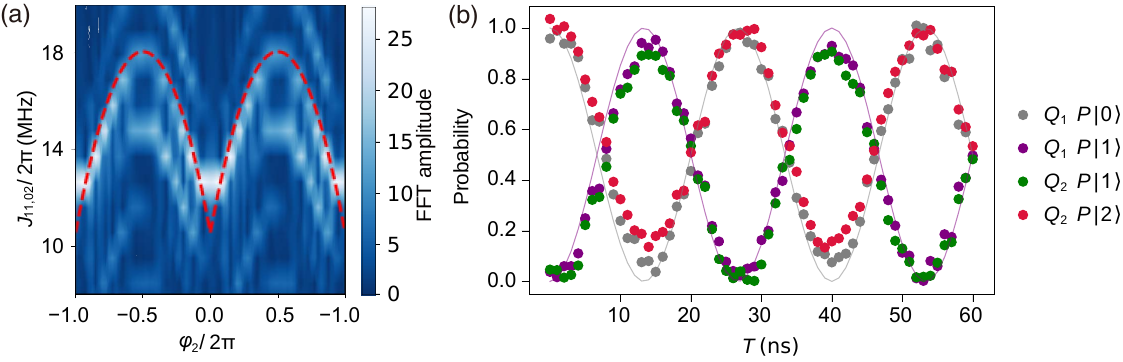}
    \caption{
        \label{paradrive_1102}
        {Modulating the phase of $|11\rangle$-$|02\rangle$ hopping. Red dash line: fitted result.}
        (a) FFT of swapping dynamics between $|11\rangle$ and $|02\rangle$ by scanning driving phase $\varphi_2$.
        (b) Swapping dynamics with evolution time $T$, observed between $|11\rangle$ and $|02\rangle$. 
    }
\end{center}
\end{figure}

The experimental result of phase modulation is shown in \rfig{paradrive_1102}(a). As an example, when we tune $\varphi_2=0$, the corresponding swapping between $|11\rangle$ and $|02\rangle$ is shown in \rfig{paradrive_1102}(b).

\section{Fock state lattice configuration}
\label{ABcage2Dto3D}

\begin{figure}[ht]
\begin{center}
    \includegraphics[width=0.7\textwidth]{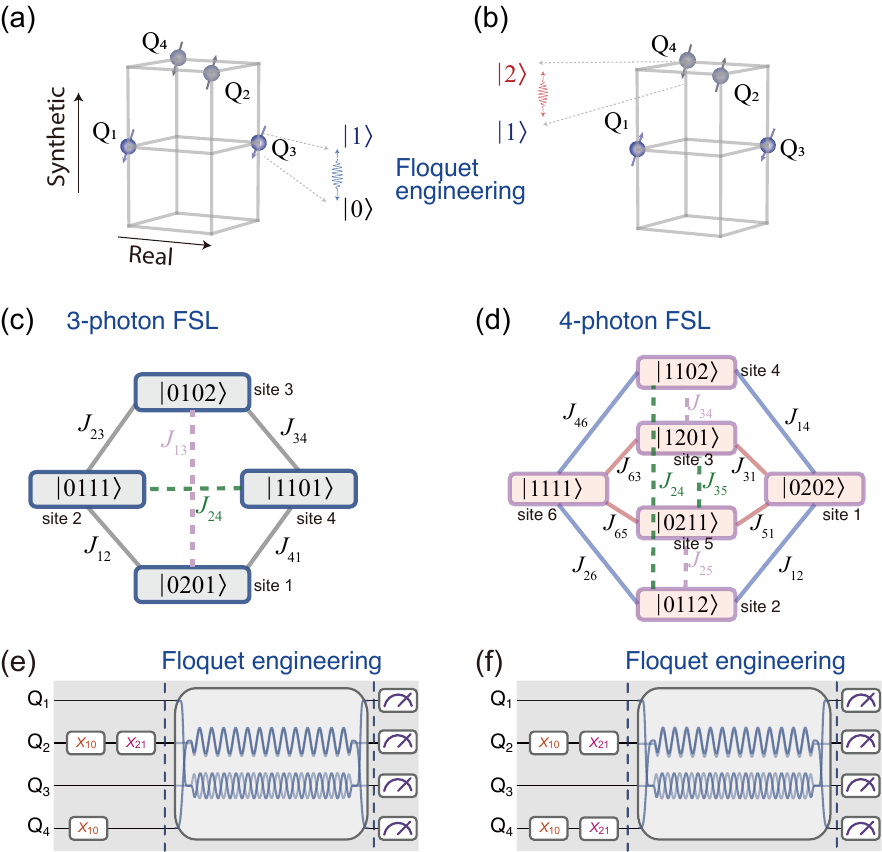}
    \caption{
        \label{schematic}
        {The schematic diagram of the synthetic Fock state lattices (FSLs).}
        (a) The rhombic platteque in real space and the FSL spanned by three-photon excitation. 
        (b) The rhombic platteque in real space and FSL spanned by four-photon excitation. 
        (c) The synthetic space configuration of four Fock states.
        (d) The synthetic space configuration of six Fock states. 
        (e) The pulse sequence for 2D AB cage.
        (f) The pulse sequence for pseudo-3D AB cage.
    }
\end{center}
\end{figure}

For a typical AB cage in real space, all sites are set in resonance. If there are two photons bounded by strong on-site interaction, they are not affected by the synthetic flux existing in the single-photon subspace. This is because the energy level of a two-photon ``doublon'' state is detuned from the non-interacting two-photon state~\cite{Martinez2023}. Whereas, in the synthetic space of this work, the interaction inherent in superconducting quantum circuit endues the small-size low-dimensional system with capability of establishing high-dimensional synthetic space. Suppose there is no interaction, if we set all qutrits in resonance, the population on any N-photon Fock state will disperse in its N-photon Fock space. Otherwise, if we detune all qutrits, the hoppings among the Fock states with the same photon number are prohibited. Either cases cannot bring the opportunity to build these FSLs.

In \rfig{schematic}(a)(b), the operational subspaces of each qutrit are marked by the black vertical arrows. We use the $\{|1\rangle,\ |2\rangle\}$ subspace instead of the $\{|0\rangle,\ |1\rangle\}$ subspace on two of the four qutrits, the ground state of this system is $|0101\rangle$. Based on this ground state, the number of blue sphere represents the extra number of excitations, the three-photon and four-photon FSLs have one and two extra excitations, respectively.

The synthetic sites for the case of three-photon FSL include ${|0201\rangle,\ |0111\rangle,\ |0102\rangle,\ |1101\rangle}$ four Fock states (\rfig{schematic}(c)), labelled as sites 1-4. The corresponding NN hoppings are labelled as $J_{12}$, $J_{23}$, $J_{34}$, and $J_{41}$, the NNN hoppings are $J_{13}$ and $J_{24}$.
The synthetic sites for the case of four-photon FSL include $|0202\rangle$, $|0112\rangle$, $|1201\rangle$, $|1102\rangle$, $|0211\rangle$, and $|1111\rangle$, labelled as sites 1-6 (\rfig{schematic}(d)). The corresponding NN hoppings are denoted by $J_{12},J_{14},J_{26},J_{46},J_{31},J_{51},J_{63}$, and $J_{65}$, the NNN hopping are $J_{24},J_{34},J_{35},J_{25}$. Where $J_{24}=J_{35}=0.5J_{25}=0.5J_{34}$ because $J_{24}$ and $J_{35}$ are formed by NNN $|01\rangle-|10\rangle$ coupling, while $J_{25}$ and $J_{34}$ are formed by NNN $|11\rangle-|02\rangle$ coupling. From this perspective, the region enclosed by these NNN hoppings is ``skewed".

The quantum circuit diagram for constructing the three-photon and four-photon FSLs are shown in \rfigs{schematic} (e-f), respectively. The parametric driving pulse on $Q_3$ is attached with a relative $\pi$ phase to other three pulses. For the free quantum walk scenario, the parametric-modulated effective NN hopping strengths are tuned to $(18.4\pm 0.4) ~\mathrm{MHz}$ in the 2D FSL, $(19.8\pm 0.6) ~\mathrm{MHz}$ in the pseudo-3D FSL. For the localization scenario, they are $(17.2\pm 0.4)$ MHz in the 2D FSL, and $(11.3\pm 0.5) ~\mathrm{MHz}$ in the pseudo-3D FSL.

\subsection{Lattice configuration for 2D AB cage}

\begin{figure}[h]
\begin{center}
    \includegraphics[width=0.7\textwidth]{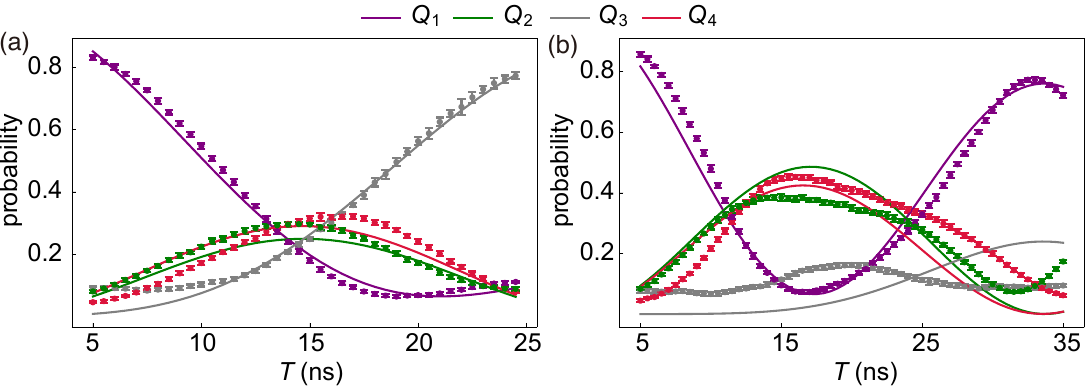}
    \caption{
        \label{resonance_drive01_tune}
        {Calibration of four-qutrit interferences in real space. Dots: experimental data. Solid curves: numerical simulation based on actual device.}
        (a) Enlarged view of the constructive interference in a four-site plaquette. The evolution starts from $Q_1$ and transmits through two paths: $Q_1-Q_2-Q_3$ and $Q_1-Q_4-Q_3$ with the same phases.
        (b) Enlarged view of the destructive interference in a four-site plaquette. The evolution starts at $Q_1$, transmits through two path: $Q_1-Q_2-Q_3$ and $Q_1-Q_4-Q_3$ with a $\pi$ phase difference.
    }
\end{center}
\end{figure}

\begin{figure}[htp]
\begin{center}
    \includegraphics[width=0.8\textwidth]{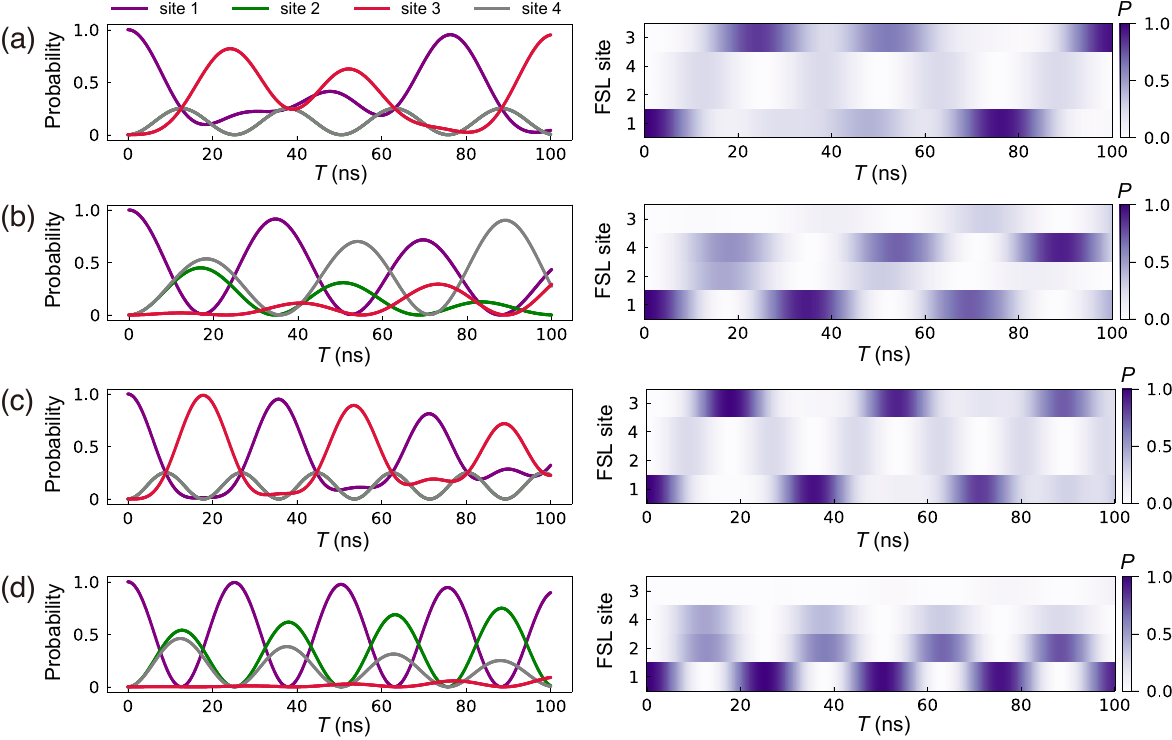}
    \caption{
        \label{2D ABcage numerical}
        {Numerical analysis of the dynamical feature of 2D AB cage.}
        (a) Zero-flux dynamics in real space.
        (b) $\pi$-flux dynamics in real space.
        (c) Zero-flux dynamics in synthetic FSL.
        (d) $\pi$-flux dynamics in synthetic FSL.
    }
\end{center}
\end{figure}

The typical AB cage model requires only NN hoppings, the NNN hoppings cause the deviation between the ideal model and our actual device. Especially when we regard actual qutrits as the lattice sites, the strong NNN hoppings lead to non negligible impact when four qutrits are in resonance (see \rfig{resonance_drive01_tune}).
Besides the NNN hoppings, numerical simulation implies that a disorder on the hopping strength $\delta J_{\text{01,10}}/2\pi \sim0.4\ ~\mathrm{MHz}$ between different legs of the plaquette leads to the mismatching of two interference paths. Consequently, if we implement the AB cage in real space,the coherent constructive and destructive patterns are maintained only in the first time-evolution period (see \rfigs{2D ABcage numerical} (a) and (b)). However, we can use the synthetic FSL sites to implement the AB cage when the qutrits are detuned. The impact of NNN hoppings are weakened in this detuned scenario (see \rfigs{2D ABcage numerical} (c) and (d)). This becomes an advantage of using the $|11\rangle-|02\rangle$ interaction to form the hopping among FSL sites.

Looking into the 2D plaquette, the total flux is captured by: $\phi = \phi_{12} + \phi_{23} + \phi_{34} + \phi_{41}$, where $\phi_{12} = \delta \varphi_{32}$, $\phi_{23} = \delta \varphi_{34}$, $\phi_{34} = \delta \varphi_{14}$, $\phi_{41} = \delta \varphi_{12}$. Therefore, $\phi=0$ can be achieved by setting $\varphi_1=\varphi_2=\varphi_3=\varphi_4=0$, $\phi=\pi$ has more than one solutions. For example, $\varphi_3=\pi/2$, $\varphi_1=\varphi_2=\varphi_4=0$.

When all driving phases are aligned to $0$, there is an effective avoided-crossing splitting between $|0102\rangle$ and $|0201\rangle$. When the phase on one leg is changed to $\pi$, the diagonal sites: $|0201\rangle$ and $|0102\rangle$ are degenerate, $|0111\rangle$ and $|1101\rangle$ are degenerate, the indirect effective hopping between opposite sites are prohibited, despite the hopping between NN sites exists. 

\subsection{Lattice configuration for pseudo-3D AB cage}
\begin{figure}[t]
\begin{center}
    \includegraphics[width=0.8\textwidth]{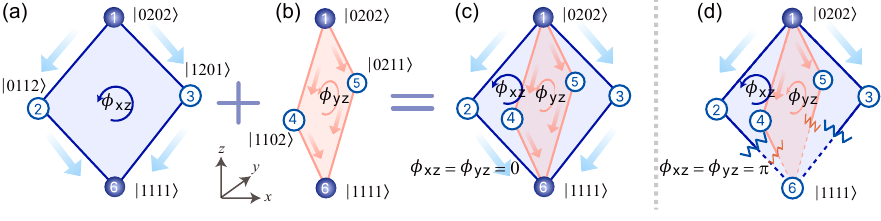}
    \caption{
        \label{schematic_2D}
        Construction of a pseudo-3D AB cage from two 2D cages.
        (a) Configuration of $xz$ plaquette.
        (b) Configuration of $yz$ plaquette.
        (c) and (d) Schematic diagrams of the pseudo-3D AB cage corresponding to flux values of 0 and $\pi$, respectively.
    }
\end{center}
\end{figure}

\begin{figure}[htbp]
\begin{center}
    \includegraphics[width=0.8\textwidth]{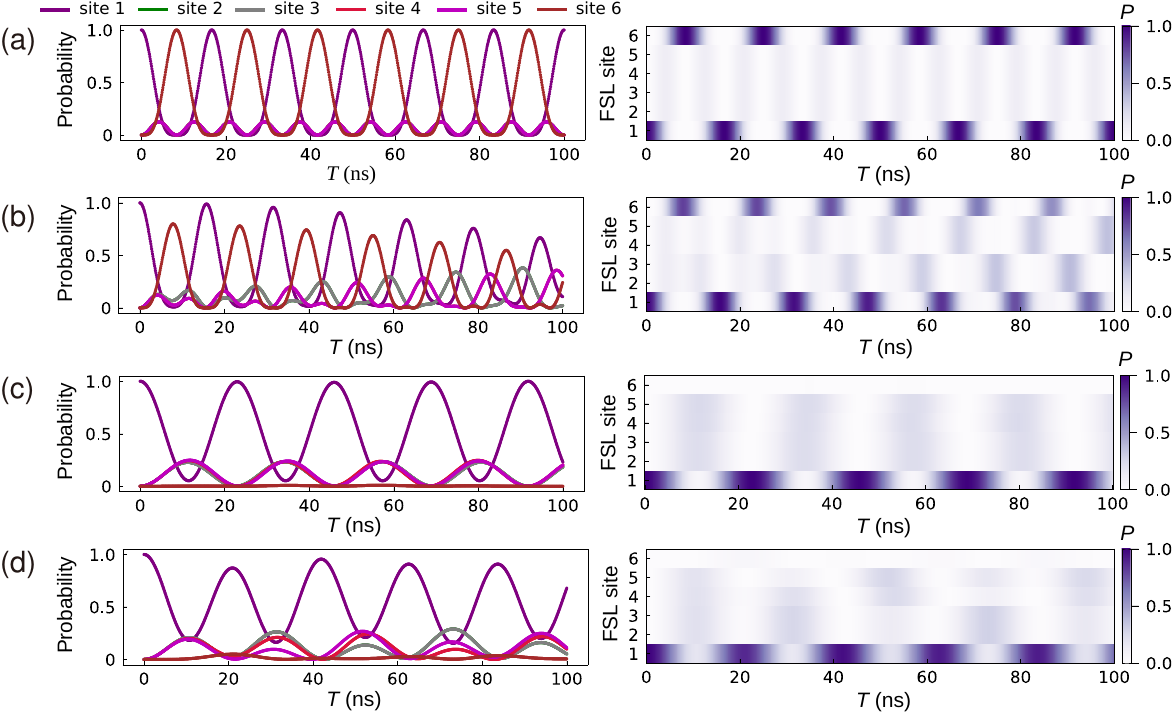}
    \caption{
        \label{num_25D_0pi}
        Numerical analysis of dynamics in the pseudo-3D AB cage.
        (a) Ideal dynamics of the 0-flux double-loop coherent superposition.
        (b) Dynamics of the 0-flux double-loop coherent superposition with a 10 MHz detuning between the two loops. Disorder introduces differences in the interference patterns of the two loops.
        (c) Ideal dynamics of the $\pi$-flux double-loop caging.
        (d) Dynamics of the $\pi$-flux double-loop caging with a 10 MHz detuning between the two loops. Disorder causes mismatches in the population oscillations between the two loops. $T$ represents the evolution time. 
    }
\end{center}
\end{figure}

The dynamical features of the pseudo-three-dimensional (3D) AB cage is illustrated in \rfig{num_25D_0pi}. We construct two interference loops: site 1, 2, 6, 3, constituting the $xz$ plaquette and site 1, 4, 6, 5, constituting the $yz$ plaquette, as shown in \rfig{schematic_2D}. The flux in $xz$ plaquette is $\phi_{xz} = \phi_{12}+\phi_{26}+\phi_{63}+\phi_{31}$, the flux in $yz$ plaquette is $\phi_{yz} = \phi_{14} + \phi_{46} + \phi_{65} + \phi_{51}$. Where $\phi_{12} = \phi_{63} = \delta\varphi_{32}$, $\phi_{26} = \phi_{31} = \delta\varphi_{14}$, $\phi_{14} = \phi_{65} = \delta\varphi_{12}$, $\phi_{46} = \phi_{51} = \delta\varphi_{34}$. The zero flux in both plaquettes is achieved by setting $\varphi_1=\varphi_2=\varphi_3=\varphi_4=0$, the $\pi$ flux is achieved by setting $\varphi_1=\varphi_2=\varphi_4=0$, $\varphi_3=\pi/2$. 
The ideal double-loop coherent interference emerges when six sites are in resonance (see \rfigs{num_25D_0pi}(a) and (c)). However, when the experimental imperfections exist, including detuning among the FSL sites, disorder on the hopping strength and phase inaccuracy, the transmission in the two loops becomes desynchronized. For example, when site 2,3 are $10$ MHz detuned from site $4,5$, the wave packet arrives at site 6 earlier from $yz$ plaquette, then it interferes with the later-arriving wave, resulting in multiple interference patterns (see \rfigs{num_25D_0pi}(b), (d) and the main text).

\subsection{Lattice configuration for 3D AB cage}

\begin{figure}[htbp]
\begin{center}
    \includegraphics[width=0.9\textwidth]{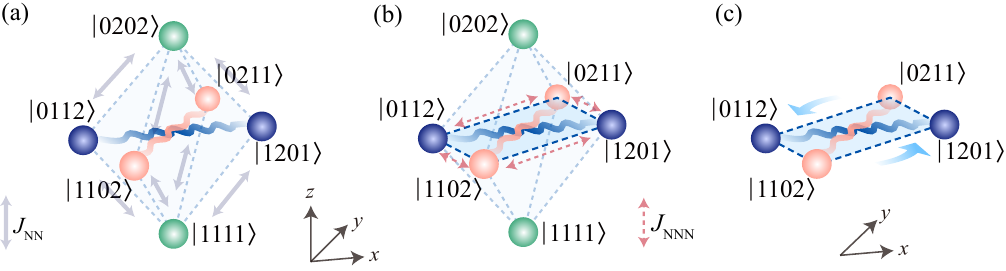}
    \caption{
        \label{schematic_3D}
        {The configuration of the hoppings in the 3D AB cage.}
        (a) Schematic of the synthetic FSL with only NN hoppings.
        (b) Configuration of the 3D AB cage comprising a synthetic FSL with NNN hoppings.
        (c) Schematic diagram of the $xy$ plaquette of the 3D AB cage.
    }
\end{center}
\end{figure}

\begin{figure}[h]
\begin{center}
    \includegraphics[width=0.8\textwidth]{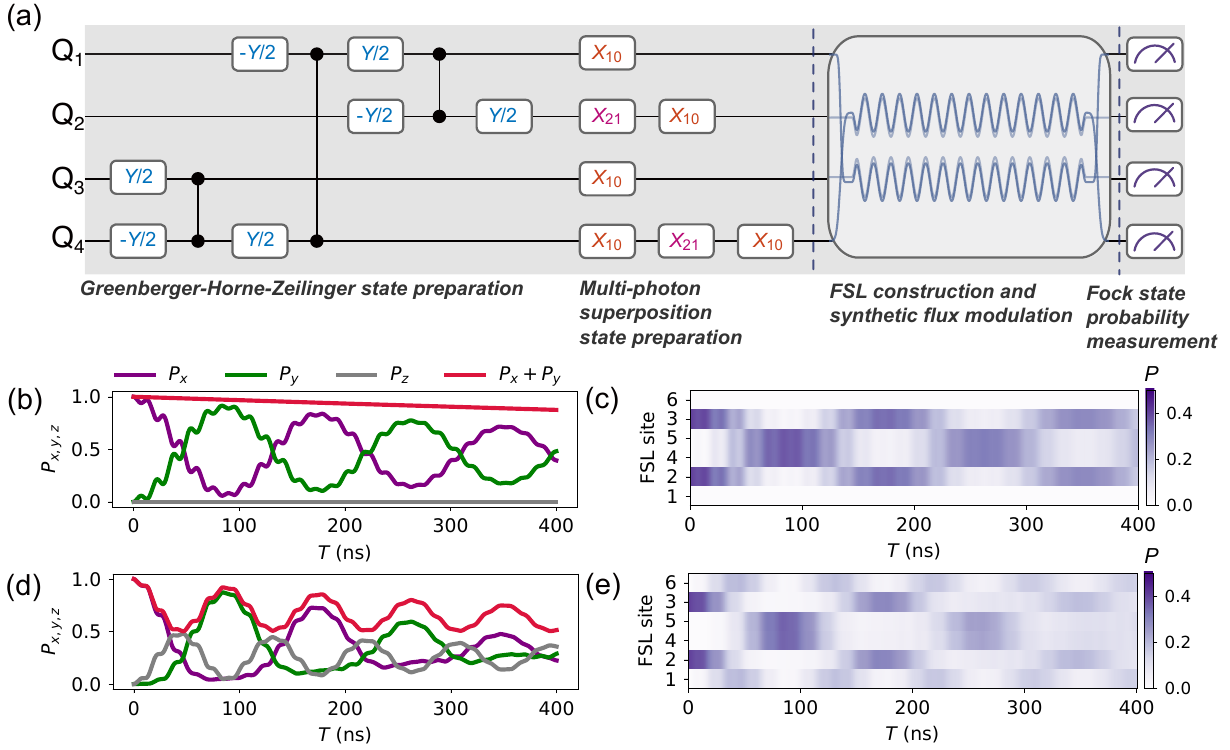}
    \caption{
        \label{numerical_3D_xy}
        {The circuit implement and numerical anlysis of the 3D AB cage.}
        (a) Quantum circuit used to synthesize the 3D AB cage.
        (b),(c) Caging dynamics with only effective NNN hopping strength $J_\mathrm{NNN}/2\pi\ \sim 2.85$ MHz.
        (d),(e) Caging dynamics with only effective NN hopping strength $J_\mathrm{NN}/2\pi\ \sim 2.85$ MHz.
    }
\end{center}
\end{figure}

The 3D AB cage demonstrated in this work arises from the combined contributions of NN (see \rfig{schematic_3D}(a)) and NNN hopping (see \rfigs{schematic_3D}(b) and (c)). Specifically, the NNN hoppings on the ``$xy$ plaquette" induces SWAP-like dynamics between the $x$ comonent (blue spheres and wavy lines) and $y$ component (yellow spheres and wavy lines) of the superposition state (see \rfig{schematic_3D}(c)). Through careful adjustments, we set the phases on the four legs of the $xz$ plaquette to be $\phi_{12} = \phi_{36} = \pi$, $\phi_{26} = \phi_{13} = 0$, and the phases on the four legs of the $yz$ plaquette to be $\phi_{14} = \phi_{56} = \pi$, $\phi_{15} = \phi_{46} = 0$. This is achieved by aligning phases on $Q_1\sim Q_4$ to $\varphi_1 = \varphi_2 = \varphi_3 = 0$, $\varphi_4 = \pi$. This results in coherent cancellation of occupation on both site 1 and site 6 (see the main text). 
The numerical simulations confirm the mechanism of this observation. As shown in \rfig{numerical_3D_xy}, panels (b-c) depict the dynamical evolution considering only NNN hoppings, and the NN hoppings in $xz$ and $yz$ plaquettes are extremely cancelled. Panels (d-e) gives a trivial case, also exhibiting an oscillation between x and y components. Where the entrapped flux inside $xz$ and $yz$ plaquettes are zero, the flat bottom region indicates that the NN hoppings dominate the exchanging between $P_x$ and $P_y$ across $P_z$. In this case, there is no caging effect on the 3D FSL. The results demonstrate that our caging pattern differs from the scenario of pure NN interference discussed in typical AB cages, since the $xy$ plaquette is completely built by NNN hoppings.

The operations on four qutrits to perform the localization phenomenon on the $xy$ plaquette are shown in \rfig{numerical_3D_xy}(a). Firstly, we prepare $Q_3$ on superposition state $\frac{1}{\sqrt{2}}(|0\rangle+|1\rangle)$, then we apply CNOT gates (comprising of $\{-Y/2, \mathrm{CZ}, Y/2\}$ gate set) in order of $Q_3$--$Q_4$, $Q_4$--$Q_1$, $Q_1$--$Q_2$ to generate a Greenberger-Horne-Zeilinger (GHZ) state: $\frac{1}{\sqrt{2}}(|0000\rangle + |1111\rangle)$, and achieve a fidelity over $0.9$. Secondly, we apply $X_{10}$ and $X_{21}$ gates to prepare state $\frac{1}{\sqrt{2}}(|0112\rangle+|1201\rangle)$, the superposition of site 2 and site 3 in the 3D FSL. These process reach a state fidelity over $0.8$. Thirdly, we quench the four qutrits to the operational energy level structure, with fast-bias flux pulses, and immediately apply parametric drive pulses on four qutrits, results in $\pi$ flux trapped inside two triangular regions. Considering the stray terms in the interaction Hamiltonian $H_{1,2}^{\mathrm{eff}}$ among those sites, the actual energy levels are slightly shifted from the ideal model. For this reason, we carefully tune the operational frequencies to ensure six sites in resonance, until the SWAP-like dynamics is localized in the $xy$ plaquette.

\section{Scaling up multi-dimensional FSL in one-dimensional loops}

\subsection{The FSL configuration with $N$ qutrits}

\begin{figure}[h]
        \begin{center}
        	\includegraphics[width=1.0\textwidth]{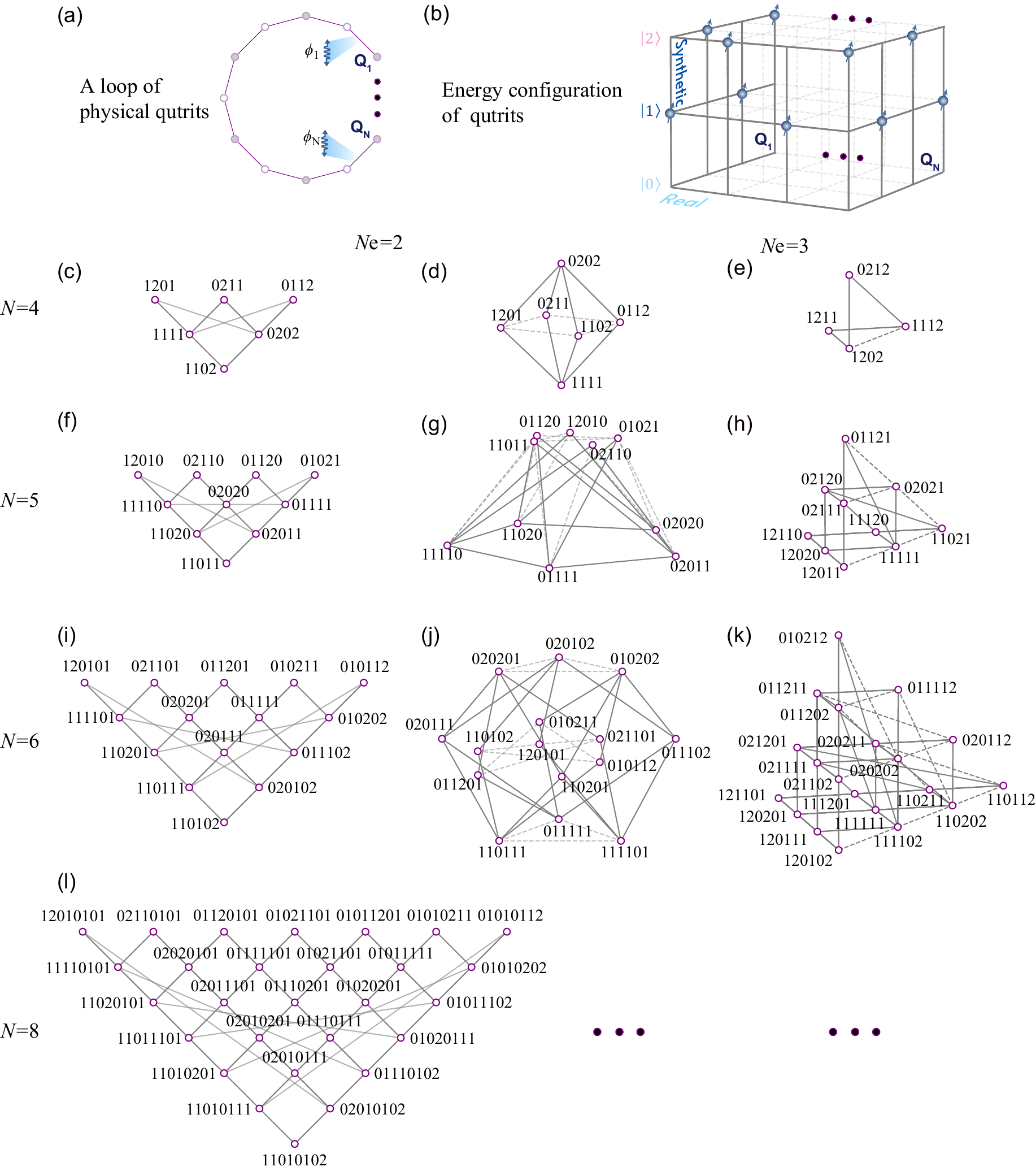}
        	\caption{
            \label{scalingqutrit}
            {Scaling-up the multi-dimensional FSLs with a 1D loop of qutrits. (a) A schematic diagram of the physical structure. Each circle represents one qutrit, the qutrits in the same color are tuned in resonance. The edges denote coupling strengths with phases: $\{\phi_1, ...\phi_{\rm{N}}\}$. (b) The energy configuration of qutrits illustrated in (a). (c,f,i,l) A planar representations of the two-excitation ($N_{\rm{e}}=2$) Fock states (purple circles) and the NN coupling (grey lines) for $N=4,5,6,8$  qutrits. (d,g,j) 3D representations of the two-excitation FSL. (d,g,j) 3D representations of the three-excitation ($N_{\rm{e}}=3$) FSL.
            } 
            }
        \end{center}
    \end{figure}

The FSL configuration is scalable to much more complicated cases. For simplicity, we consider the case of a loop of qutrits (see \rfig{scalingqutrit}{(a)}). By arranging the working frequencies of the qutrits, $Q_1$ to $Q_N$, in a staggered configuration (as shown in \rfig{scalingqutrit}{(b)}), the system forms a platform with internal dimensions corresponding to the multi-excitation Hilbert space.
The ground state of this system is defined as $|0101...01\rangle$, when the total number of excitations is $N_{\rm{e}}=2$, the FSL configurations for $N=4,5,6$ and $8$ qutrits can be illustrated as \rfigs{scalingqutrit}{(c,f,i,l)}. Assuming uniform NN coupling strengths and the inclusion of NNN couplings, these FSLs can be projected into 3D space, as shown in \rfigs{scalingqutrit}{(d,g,j)} (the 3D structure for $N=8$ is omitted due to its complexity).
When $N_{\rm{e}}=3$, the synthetic FSLs exhibit multi-layer structures of the $N_{\rm{e}}=2$ FSLs, as illustrated in \rfigs{scalingqutrit}{(e,h,k)}.  
For convenience, we relabel the site number via the indexes of the excited qutrits. For $N=6$, $N_{\rm{e}}=2$, the excited qutrits are indexed as $(i,j)$, where $i=1,2,...,N-1,\ j=i+1,i+2,...,N$, hence Fock states $|120101\rangle...|010112\rangle$ can be relabeled as $\sum_{x=1}^{i-1}(N-x) + (j-i)$, namely FSL sites $1\sim15$. For $N=6$, $N_{\rm{e}}=3$, the excited qutrits are $(i,j,k)$, where $i=1,2,...,N-2,\ j=i+1,i+2,...,N-1,\ k=j+1,j+2,...,N$, Fock states $|121101\rangle...|010212\rangle$ are then relabeled as $\sum_{x=1}^{i-1}\sum_{y=x+1}^{N}(N-y) + \sum_{y=i+1}^{j-1}(N-y) + (k-j)$, namely FSL sites $1\sim20$. 

\clearpage

\subsection{Demonstration of FSL with $N=6$ and $N_{\rm{e}}=2$}

\subsubsection{Device information}
\begin{figure}[h]
        \begin{center}
            \includegraphics[width=1.0\textwidth]{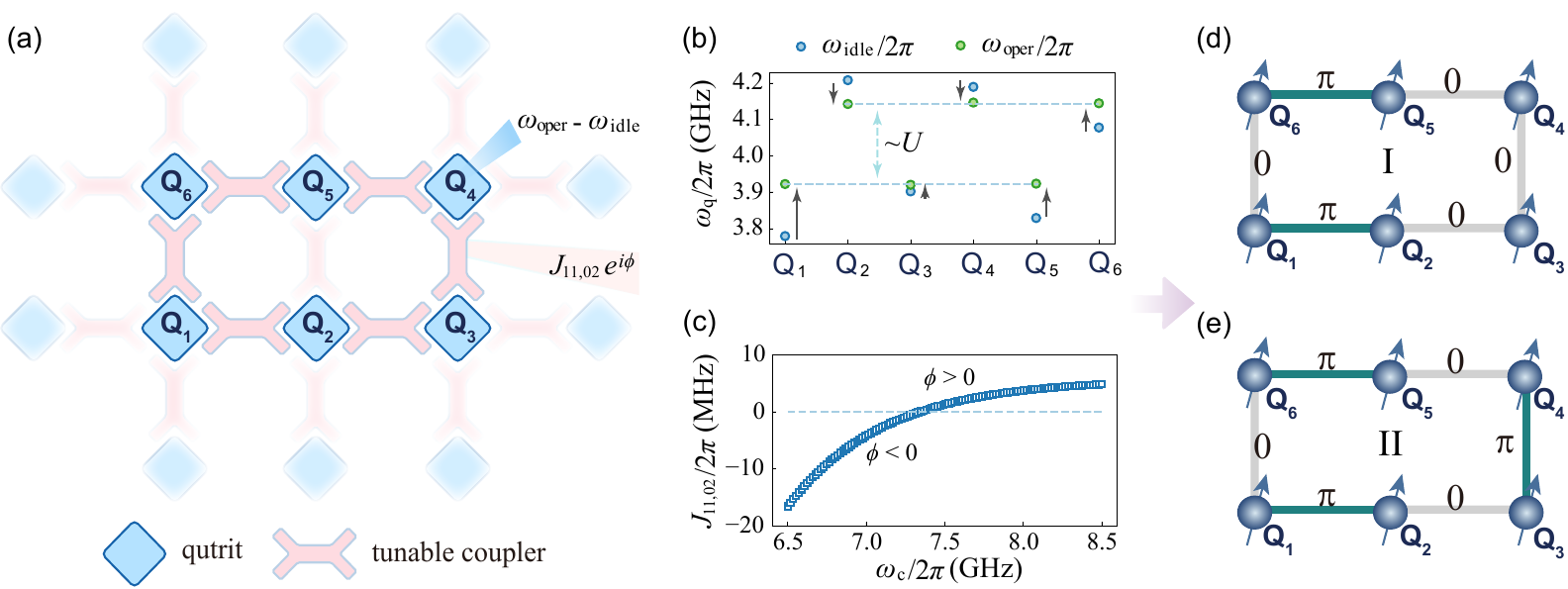}
        	\caption{
            \label{processor6q}
            {(a) Schematic of the quantum processor, highlighting the qutrits and tunable couplers used in this work (solid components). (b) Frequency distribution of the qutrits employed in the experiment. (c) The coupling strength $J_{\rm{11,02}}$ as a function of the estimated coupler frequency $\omega_{\rm{c}}$. (d) The flux configuration I used in the FSL mode. (e) The flux configuration II.} 
            }
        \end{center}
    \end{figure}

\begin{table}[t]
    \centering
    \caption{6-qutrit device information}
    \label{qutrit_chip2_info}

    \begin{tabular}{|p{4cm}<{\centering}|p{1.5cm}<{\centering}|p{1.5cm}<{\centering}|p{1.5cm}<{\centering}|p{1.5cm}<{\centering}|p{1.5cm}<{\centering}|p{1.5cm}<{\centering}|}
        \hline
	\hline	        
        \text{Qutrit} & $Q_1$ & $Q_2$  & $Q_3$  & $Q_4$  & $Q_5$ & $Q_6$\\ \hline
        \text{$\omega_\mathrm{idle}/2\pi$ (GHz)} & 3.779 & 4.208 & 3.902 & 4.189 & 3.829 & 4.077 \\ \hline
        \text{$\omega_\mathrm{oper}/2\pi$ (GHz)} & 3.922 & 4.142 & 3.920 & 4.146 & 3.923 & 4.144 \\ \hline
        \text{$\omega_r/2\pi$ (GHz)} & 6.161 & 6.121 & 6.212 & 6.074 & 6.218 & 6.113 \\ \hline
        \text{$U_i/2\pi$ (MHz)} & -224.7 & -214.8 & -227.4 & -222.7 & -224.0 & -217.1 \\ \hline
        \text{$F_{00}$} & 0.788 & 0.824 & 0.857 & 0.889 & 0.877 & 0.758 \\ \hline
        \text{$F_{11}$} & 0.756 & 0.719 & 0.829 & 0.657 & 0.758 & 0.767 \\ \hline
        \text{$F_{22}$} & 0.774 & 0.813 & 0.837 & 0.777 & 0.868 & 0.759\\ \hline
        \text{$T_1^{10} (\mu s)$} & 89.8 & 68.7 & 57.9 & 66.5 & 62.1 & 89.4 \\ \hline
        \text{$T_2^{10}\ (\mu s)$} & 4.62 & 4.08 & 1.99 & 3.43 & 3.38 & 3.77\\ \hline
        \textbf{$T_1^{21}\ (\mu s)$} & 45.9 & 60.8 & 30.9 & 43.5 & 45.6 & 64.2 \\ \hline
        \textbf{$T_2^{21}\ (\mu s)$} & 1.63 & 1.83 & 1.56 & 2.71 & 1.43 & 2.76 \\ \hline

    \end{tabular}
\end{table}

To further illustrate the scalability and generality of the approach, we extend the configuration to a six-qutrit loop to construct the FSL for $N = 6$, exploring its phase modulation and coherent dynamics.
The quantum processor is schematically illustrated in \rfig{processor6q}{(a)}, comprising qutrits arranged in a 2D square lattice with couplings mediated by tunable couplers. For this experiment, we select six qutrits, $Q_1$ to $Q_6$, forming a closed-loop configuration to emulate the $N$-qutrit FSL. The Hamiltonian of this simulator can be expressed as ($N = 6$ here):
\begin{equation}\label{Hami}
   H/\hbar= \sum_{i=1}^{N}\!\left[\omega_{i}a_{i}^{\dagger}a_{i}\!+\!\frac{1}{2}U_{i}n_{i}\left(n_{i}\!-\!1\right)\right]\! +\!\sum_{i<j}g_{ij}(a_{i}^{\dagger}a_{j}\!+\!a_{j}^{\dagger}a_{i}),
\end{equation}
where $a_{i}$ and $n_{i}=a_{i}^{\dagger}a_{i}$ denote the bosonic annihilation and particle number operators. 
Initially, the qutrits are prepared in their respective states and set to idle frequencies, $\omega_{\rm{idle}}$. Subsequently, all qutrits are simultaneously quenched to their operating frequencies, $\omega_{\rm{oper}}$, using flux pulses with amplitudes determined by $\omega_{\rm{oper}} - \omega_{\rm{idle}}$ (see \rfigs{processor6q}{(a-b)} and \rtab{qutrit_chip2_info} for detailed parameters).
The tunable couplers between the nearest-neighbor qutrits allow the coupling strength $J_{\rm{11,02}}$ to vary between positive and negative values, corresponding to $\phi > 0$ and $\phi < 0$ in the model (see \rfig{processor6q}{(c)}). 
Since the couplers' design bypasses the readout resonators, their frequencies $\omega_{\rm{c}}$ are estimated via numerical simulations. To explore flux-controlled interference within the FSL, we implement two distinct flux configurations, illustrated in \rfigs{processor6q}{(d-e)}.
In configuration I, the coupling strengths between $Q_2$ and $Q_3$, $Q_3$ and $Q_4$, $Q_4$ and $Q_5$, and $Q_6$ and $Q_1$ are tuned to $J_{23} = J_{34} = J_{45} = J_{61} = +J/2\pi = 5.0~\mathrm{MHz}$, while the couplings between $Q_1$ and $Q_2$, and $Q_5$ and $Q_6$, are set to $J_{12} = J_{56} = -J/2\pi = -5.0~\mathrm{MHz}$. 
In configuration II, the coupling strength between $Q_3$ and $Q_4$ is reversed to $-J/2\pi$. Leveraging the rotational symmetry of the loop, edges with $+J/2\pi$ are defined as $\phi = 0$, and those with $-J/2\pi$ as $\phi = \pi$.
This approach demonstrates a generalized method for constructing multi-dimensional FSLs, independent of specific chip designs. The requirement for a 2D processor with tunable couplers is not stringent, offering an opportunity to reconsider experimental costs when studying multi-dimensional quantum systems.
In previous sections, we have described the physical structure of the $N=4$ system as a plaquette, with the synthetic space for $N_{\rm{e}}=2$ forming an octahedron.
This structure serves as a building block for the 3D AB cage.
For example, we can fold two edges of the 6-qutrit loop in three ways ({see \rfigs{processor6q}(d-e)}): ($Q_1$ - $Q_2$, $Q_4$ - $Q_5$), ($Q_2$ - $Q_3$, $Q_5$ - $Q_6$), and ($Q_3$ - $Q_4$, $Q_6$ - $Q_1$), therefore it returns to the 4-qutrit plaquette. Each folded loop configuration shares two qutrits with another configuration, indicating that the FSL is constructed from three nested octahedra, each sharing two sites. These shared sites split into pairs: $|110102\rangle$ - $|011201\rangle$, $|021101\rangle$ - $|010112\rangle$ and $|010211\rangle$ - $|120101\rangle$ ({as shown in \rfig{scalingqutrit}(j)}). These six sites have a coordination number of 2, while all other sites have a coordination number of 4. From this perspective, these sites serve as the ``boundary” of the multi-dimensional configuration, similar to the rhombic lattice.

\subsubsection{Mechanism of coherent interference in quantum superposition states}

\begin{figure}[h]
    \begin{center}
    \includegraphics[width=0.9\textwidth]{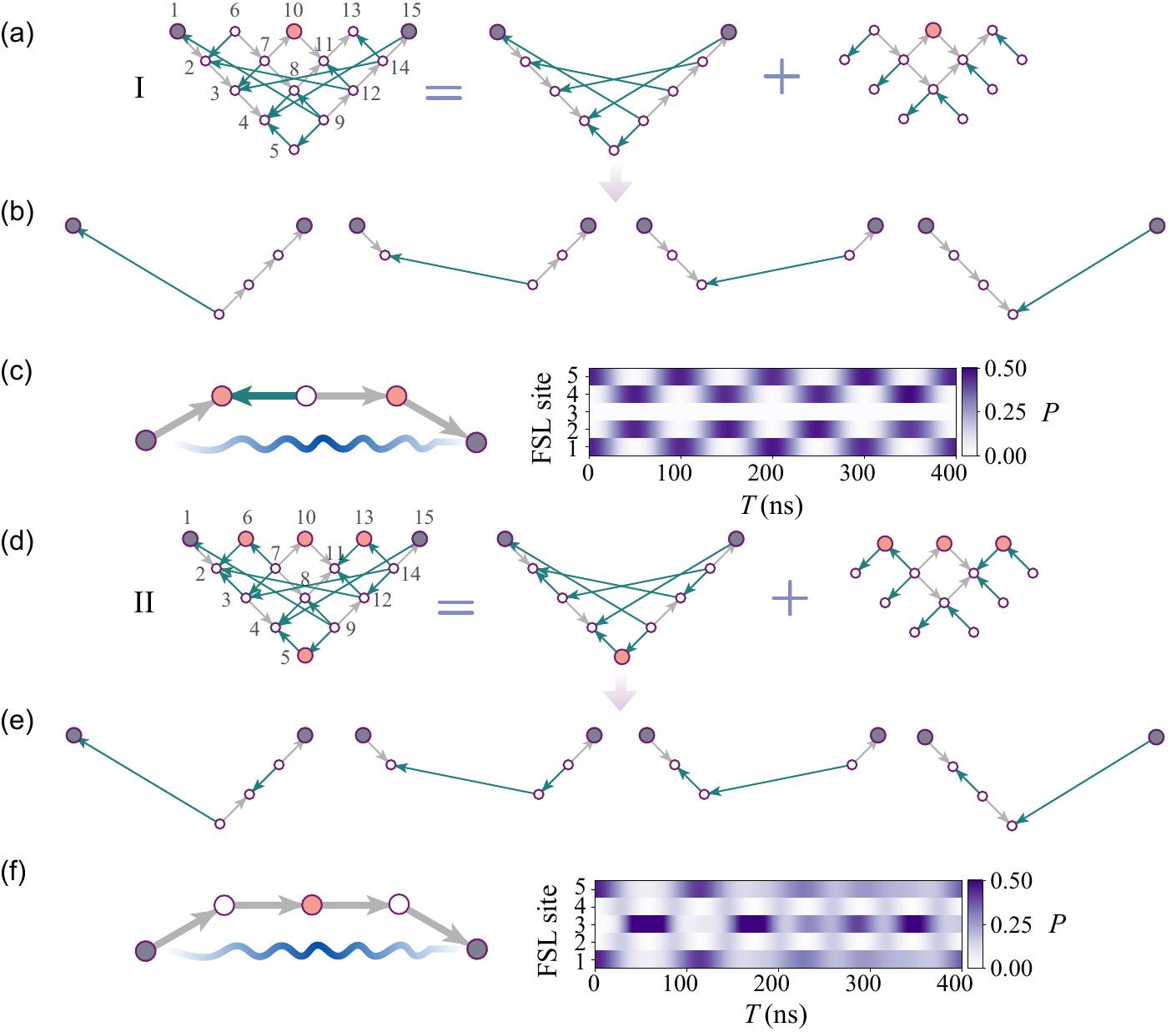}
        \caption{
        \label{principle_of_6Q_chain5q}
        {Schematic diagram for coherent interference of quantum superposition state in FSL with $N_{\rm{e}}=2$.} 
        {(a-c) The principle of interference assisted by entanglement with flux I. a-b: decomposition of the FSL configuration. c: numerical analysis of the constructive interference dynamics of a superposition state in a 1D chain.}
        {(d-f) The principle of interference assisted by entanglement with flux II. d-e: decomposition of the FSL configuration. f: numerical analysis of the destructive interference dynamics of a superposition state in a 1D chain.}
         }
    \end{center}
\end{figure}

The constructive and destructive interference paths of the superposition states in the FSL, shown in \rfig{scalingqutrit}{(i)}, can be decomposed to two parts (see \rfigs{principle_of_6Q_chain5q}{(a,d)}). The right parts for two configurations are similar to the plaquette structure with the same total flux, the left parts can be understood as the combination of 1D chains with entangled ends and edges oriented in either the positive or negative direction (see \rfigs{principle_of_6Q_chain5q}{(b,e)}).
The positive direction is defined according to the physical loop of qutrits: $Q_1 \rightarrow Q_2 \rightarrow Q_3 \rightarrow Q_4 \rightarrow Q_5 \rightarrow Q_6 \rightarrow Q_1$. For example, the edge between sites 1 and 2 is connected by the $11-02$ coupling between $Q_2$ and $Q_3$. If the phase of this coupling is 0, the edge direction is defined as positive; if the phase is $\pi$, the edge direction is defined as negative.
For $N=6$ and $N_{\rm{e}}=2$, with the initial state at sites 1 and 15, the lattice configuration of the FSL can be decomposed as shown in \rfigs{principle_of_6Q_chain5q}{(a-b)} and \rfigs{principle_of_6Q_chain5q}{(d-e)} for flux configurations I and II, respectively. The key difference between the two flux configurations lies in the interference paths between sites 1 and 15, as illustrated in \rfigs{principle_of_6Q_chain5q}{(b-c)} and \rfigs{principle_of_6Q_chain5q}{(e-f)}. 
In these 1D chains, which consist of 5 FSL sites and 4 edges, the dynamics depend on the number of negative-directional edges. When the number of negative-directional edges is even, the total phase of the chain is $2n\pi$ (with $n = 0, 1, 2, \dots$), resulting in constructive interference at the middle site, as depicted in \rfig{principle_of_6Q_chain5q}{(c)}. In this case, the population at the two ends undergoes coherent walk across the chain, leading to constructive interference at the central site. The superposition of paths adhering to this mechanism generates the pattern shown in 
\rfig{principle_of_6Q_chain5q}{(a)}.
Conversely, when the number of negative-directional edges is odd, the total phase of the chain is $(2n+1)\ \pi (n=0,1...)$, resulting in destructive interference at the middle site and constructive interference at the second and fourth sites, as shown in \rfig{principle_of_6Q_chain5q}{(f)}. This result occurs because the second and fourth sites are equidistant from both ends, taking into account the edge directionality.

\subsection{Extended data for $N=6$}

\begin{figure}[h]
\begin{center}
    \includegraphics[width=0.7\textwidth]{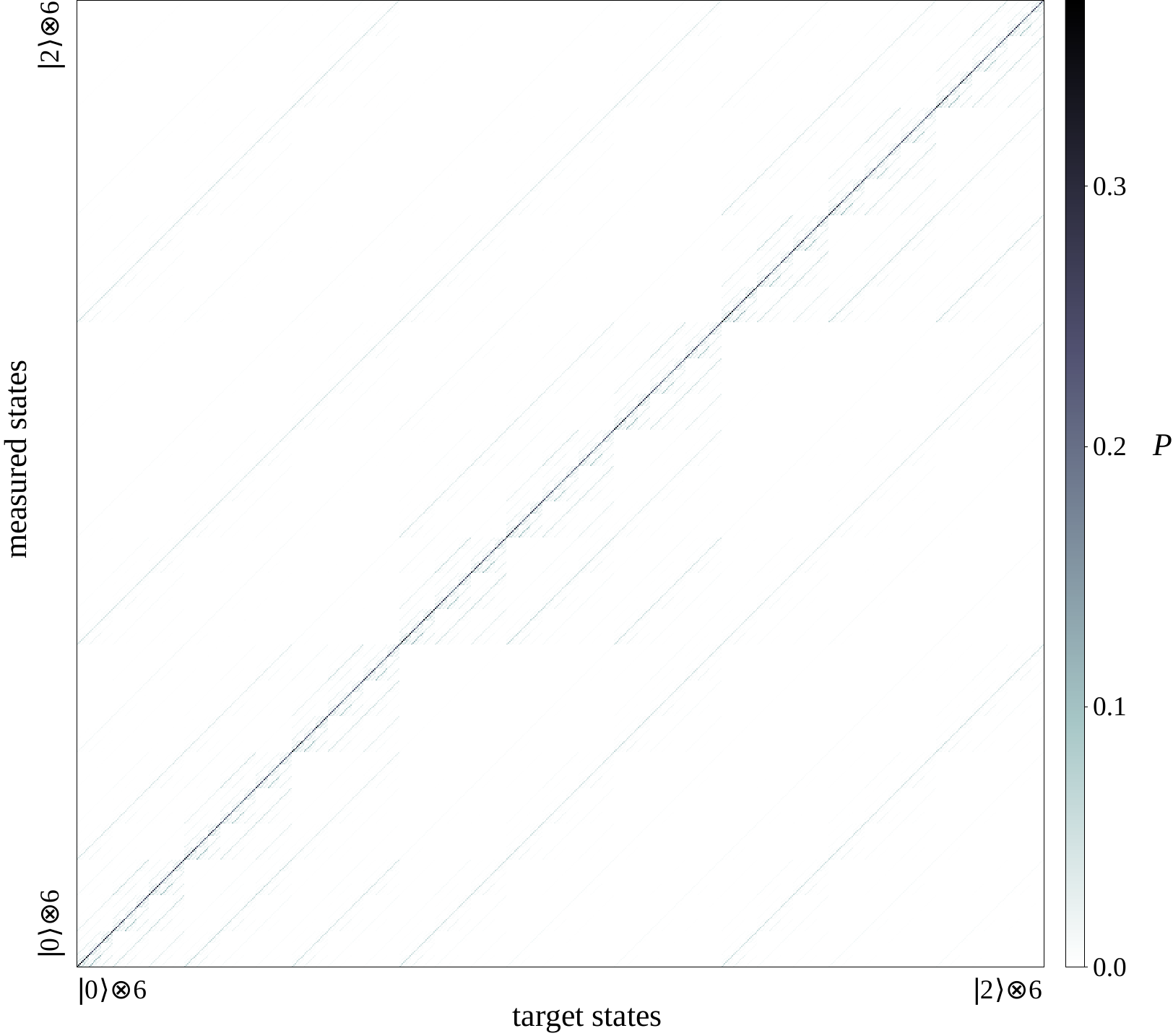}
    \caption{
        \label{ReadCrossMat_6q}
        {The readout crosstalk matrix.}
        {The six-qutrit system is prepared and measured in the computational basis spanning 729 states ($i\otimes j \otimes m \otimes n\otimes k\otimes l$, with $i,j,m,n,k,l \in\{|0\rangle, |1\rangle,|2\rangle\}$). The color bar represents the state-preparation fidelity, while off-diagonal elements correspond to crosstalk errors.}
    }
\end{center}
\end{figure}

The state-preparation efficiency is an essential requirement in synthesizing complex FSLs. In this work, we prepare and measure Fock states of the form $i\otimes j \otimes m \otimes n\otimes k\otimes l$, where $i,j,m,n,k,l \in\{|0\rangle, |1\rangle,|2\rangle\}$. The joint multi-qutrit readout parameters are optimized using the method described in Section \ref{Fockstate_SPAM}. 
The resulting data is shown in \rfig{ReadCrossMat_6q}, where the target states represent the independently prepared Fock states, and the measured states correspond to the bases in the joint measurement. To mitigate SPAM errors, the data presented in the main text and subsequent sections are processed using the inverse of this readout matrix.

\begin{figure}[hbt]
        \begin{center}
        	\includegraphics[width=1.0\textwidth]{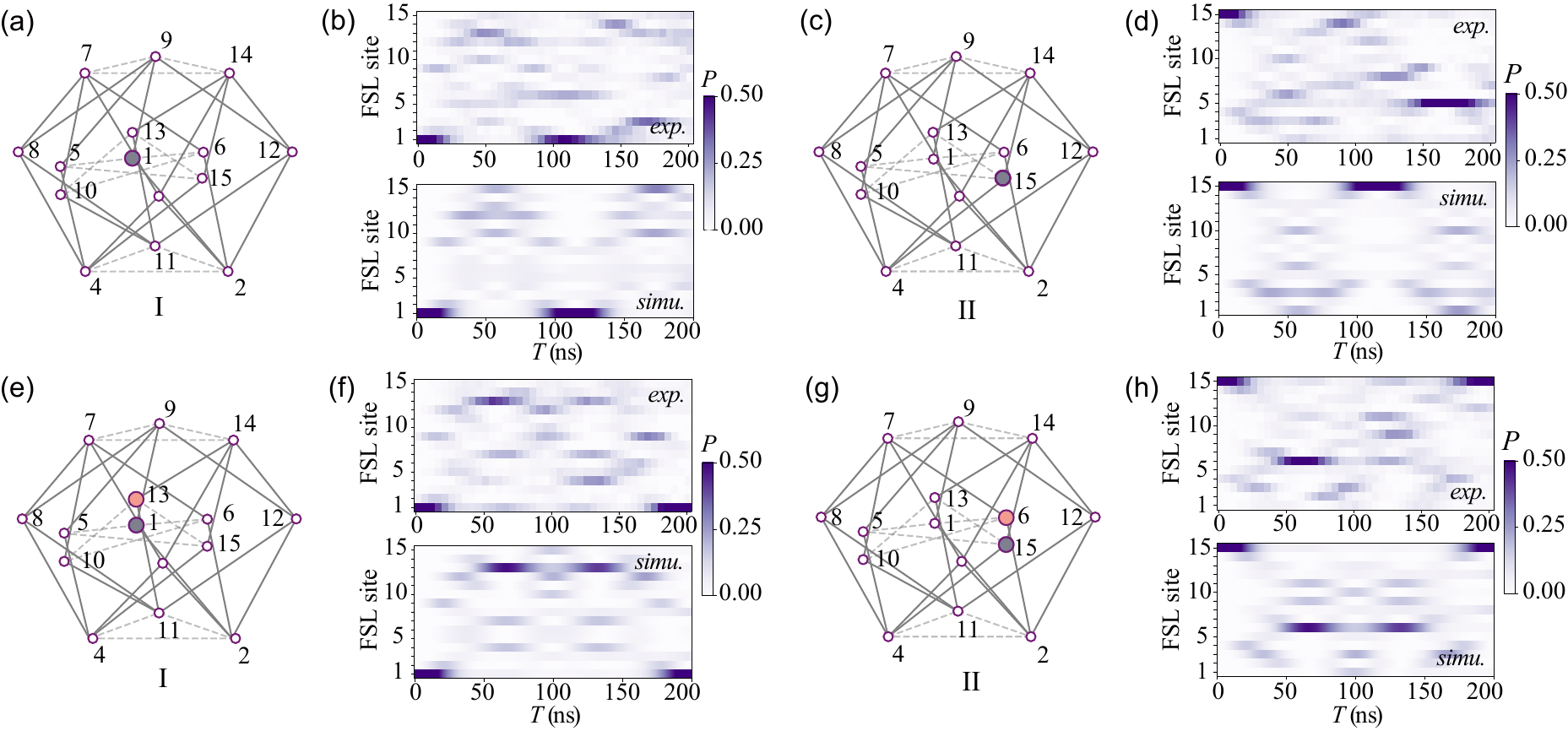}
        	\caption{
            \label{twopho_FSL_6qutrit}
            {Demonstration of the dynamics of a single bounded two-excitation state in a FSL consisting of six qutrits.} 
            {(a) FSL configuration with flux I. Left panel shows the flux modulation on the edges in real space. Right panel presents the 3D projection of the FSL. The initial state is localized at site 1 with flux configuration I.}
            {(b) Experimental (upper panel) and numerical simulation (lower panel) of the configuration shown in (a), with the initial state at site 1.}
            {(c) 3D projection of the FSL where the initial state is on site 15, and localized with flux I.}
            {(d) Dynamics of the configuration shown in (c) with the initial state at site 15.}
            {(e) FSL configuration with flux II. Left panel shows the flux modulation in real space. Right panel indicates the initial state (solid blue circle) and the site where a periodic positive superposition is observed dominantly (solid red circle).}
            {(f) Experimental (upper panel) and numerical simulation (lower panel) of the configuration in (e), with the initial state at site 1.}
            {(g) 3D projection of the FSL with the initial state at site 15 (blue circle). Periodic positive superposition is observed at site 6 (red circle).}
            {(h) Experimental (upper panel) and numerical simulation (lower panel) of the configuration in (g), with the initial state at site 15.}
            }
        \end{center}
    \end{figure}

We demonstrate the ``AB caging" phenomena in both two-excitation and three-excitation FSLs. Similar to the previous demonstration on a plaquette, the system is initialized on a single site (site 1 or site 15 in our experiment), followed by flux modulation. The photon population on different FSL sites experiences distinct interference loops, resulting in varying superpositions across sites.
When the flux is set to configuration I (see \rfigs{twopho_FSL_6qutrit}(a) and (c)), the population remains localized on initial sites 1 or 15 (see \rfigs{twopho_FSL_6qutrit}(b) and (d)). 
In flux configuration II (see \rfig{twopho_FSL_6qutrit}{(e)}), we observe a periodic positive superposition at sites different from the initial site.
Specifically, when the system starts at site 1, the dominant superposition is observed at site 13 (see \rfig{twopho_FSL_6qutrit}{(f)}), while for an initial state at site 15, the dominant superposition shifts to site 6 (see \rfig{twopho_FSL_6qutrit}{(h)}). 
The experimental results align with the corresponding numerical simulations, as shown in \rfigs{twopho_FSL_6qutrit}(f) and (h). The deviations in the experimental data come from the unexpected frequency detuning from calibration process, non-uniformity on the coupling strength between NN FSL sites, leakage error induced by residual coupling among high-energy-level states and two-level defects on our processor. These factors, both from imperfections on our hardware comparing with the ideal model, and inaccuracy in experimental process, have influence on the quality of localization and delocalization patterns in similar mechanisms as explained in Section \ref{ABcage2Dto3D}. These issues are soluable with improvement on our technical level. 

\begin{figure}[hbt]
    \begin{center}
        \includegraphics[width=1.0\textwidth]{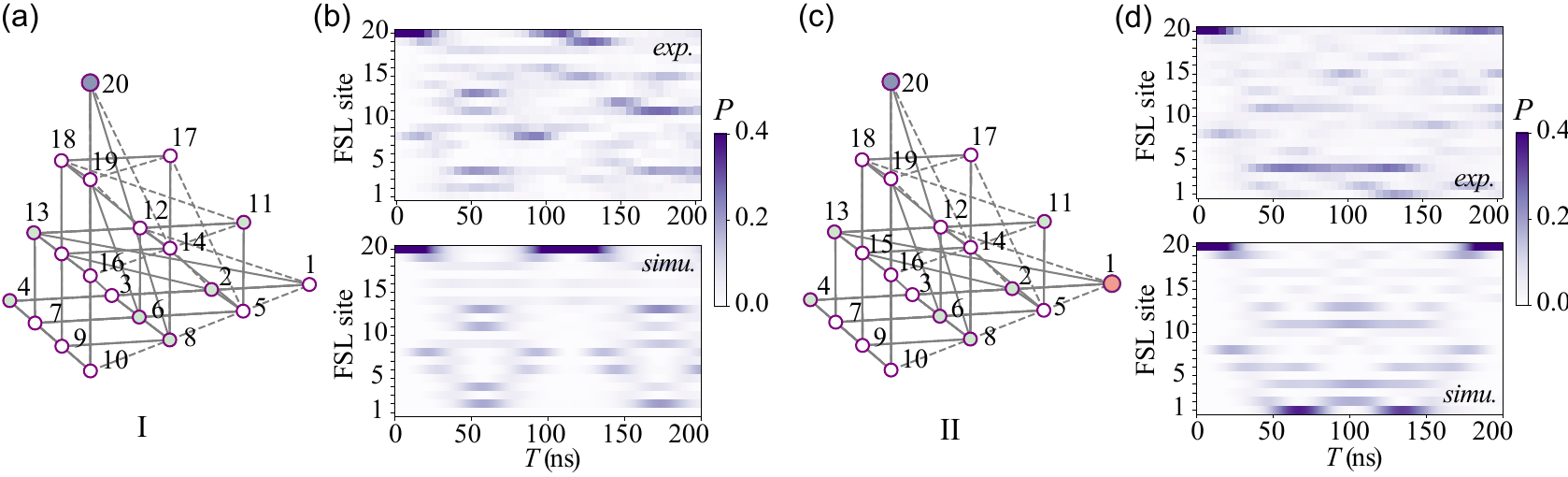}
        \caption{
        \label{Threepho_FSL_6qutrit}
        {Demonstration of the dynamics of a single bounded three-excitation state in a FSL consisting of six qutrits.} 
        {(a,c) Flux configuration (upper panel) and lattice configuration (lower panel) of the three-excitation FSL with six qutrits.}
        {(b,d) Interference dynamics in the FSL with the initial state placed on site 20: (b) Flux modulation set to configuration I. (d) Flux modulation set to configuration II.}
        }
    \end{center}
\end{figure}

For the three-excitation FSL, we initialize the system at site 20. The flux modulation again controls the dynamics across specific sites. 
A key difference between these two patterns is that in flux configuration I, the population is completely canceled at site 1 (see \rfigs{Threepho_FSL_6qutrit}{(a, b)}), while in flux configuration II, the population can reach site 1 (see \rfigs{Threepho_FSL_6qutrit}{(c, d)}).

From these data, we can conclude that the evolution of states initialized on single FSL sites of these multi-dimensional AB cages is consistent with the basic building block.

\clearpage

%